\documentclass{article} 

\usepackage{versions} 

\includeversion{full} 

\usepackage{graphicx}
\usepackage{amssymb}
\usepackage{amsmath}
\usepackage{amsthm}

\usepackage{xspace}
\usepackage{url}

\newcommand{\pow}[1]{{\cal P}(#1)}
\newcommand{\restrict}{\upharpoonright} 
\newtheorem{theorem}{Theorem}
\newtheorem{proposition}{Proposition}

\newtheorem{definition}{Definition}

\newcommand{\agts}{\mathit{Agts}}

\newcommand{\be}{\begin{enumerate}}  
\newcommand{\ee}{\end{enumerate}}  

\newcommand{\bi}{\begin{itemize}}  
\newcommand{\ei}{\end{itemize}}  

\newcommand{\Prop}{V}
\newcommand{\vars}{\mathit{vars}}

\newcommand{\rimp}{\Rightarrow}
\newcommand{\dimp}{\Leftrightarrow}

\newcommand{\assgt}{\mathit{assgt}}
\newcommand{\bisim}{R}

\newcommand{\ks}{M}

\newcommand{\vs}{\mathcal{M}}
\newcommand{\Ovars}{O}

\newcommand{\vstoks}{\mathit{ks}}
\newcommand{\kstovs}{\mathit{vs}}

\newcommand{\relss}{{\cal S}}
\newcommand{\sys}{{\cal I}}

\begin{document} 

\title{Optimizing Epistemic Model Checking using Conditional Independence\footnote{
Version of Oct 12, 2016. Work supported by US Air Force, Asia Office of Aerospace Research and Development, 
grant AFOSR FA2386-15-1-4057. 
Thanks to Xiaowei Huang and Kaile Su for some preliminary discussions and investigations
on the topic of this paper.} 
} 
\date{} 

\begin{full} 
\author{Ron van der Meyden} 
\end{full}

\maketitle

\begin{abstract} 
Conditional independence reasoning has been shown to be helpful in the 
context of Bayesian nets to optimize probabilistic inference, 
and related techniques have been applied to speed up a number of 
logical reasoning tasks in boolean logic by eliminating irrelevant parts of the formulas. 
This paper shows that conditional independence reasoning can also be 
applied to optimize epistemic model checking, in which one 
verifies that a model for a number of agents operating with 
imperfect information satisfies a formula expressed in a modal multi-agent logic of knowledge. 
An optimization technique is developed that precedes the use of a 
model checking algorithm 
with an analysis that applies conditional independence reasoning to reduce the size of the model. 
The optimization has been implemented in the epistemic model checker MCK.  
The paper reports experimental results demonstrating that it can yield multiple orders of magnitude
performance improvements. 
\end{abstract}

\section{Introduction} 

\noindent  
Epistemic model checking \cite{mck} is a technique for the verification of 
information theoretic properties, stated in terms of a modal logic of knowledge, 
in systems in which multiple agents operate with imperfect information of
their environment.  It has been applied to settings that include 
diagnosis \cite{BozzanoCGT15}, and reasoning in game-like settings \cite{HuangMM11,HuangRT13,DitmarschHMR06}, 
concurrent hardware protocols \cite{BaukusM04} and security protocols \cite{BatainehMeyden11,BoureanuCL09,MS}. 

In dealing with imperfect information, the models of epistemic model checking 
can be viewed as a discrete relative of probabilistic models. 
The Bayesian net literature has developed some very effective techniques 
for the optimization of probabilistic reasoning
based on the elimination of variables and conditional independence 
reasoning \cite{KollerFriedman,pearlprobbook}.  Similar ideas 
have been shown to be applicable to reasoning in propositional 
logic \cite{Darwiche97}. 

The contribution of the present paper is to demonstrate that these 
conditional independence techniques from the Bayesian Net literature can also be applied in 
the context of epistemic model checking. We develop a generalization of  these techniques for 
a multi-agent modal logic of knowledge, that enables
model checking computations for this logic to be optimized by reducing the
number of variables  that need to be included in data structures
used by the computation. 

In epistemic model checking, one represents the 
model as a concurrent program, in  which 
each of the agents executes a protocol in the
context of an environment. 
We provide a symbolic execution method for generating
from this concurrent program  a directed acyclic graph 
representing the model using symbolic values. 
Conditional independence reasoning is used to 
reduce this directed graph to a smaller one that 
removes variables that can be determined to be irrelevant to the 
formula to be model checked. 
Epistemic model checking can then be performed in this
reduced representation of the model using any of a number of 
approaches, including binary decision diagrams \cite{BCMDH90} and 
SAT-based techniques (bounded model checking \cite{BiereCCSZ03}). 

We have implemented the technique in the epistemic model checker MCK \cite{mck}. 
The technique developed can be applied for other semantics and algorithms, but 
we focus here on agents with \emph{synchronous perfect recall} and model check the 
reduced representation using binary decision diagram techniques. 
The synchronous perfect recall semantics presents
the most significant challenges to the computational cost of 
epistemic model checking, since it 
leads to a rapid blowup in the number of variables that need
to be handled by the symbolic model checking algorithms. 

The paper presents experimental results that demonstrate 
that the conditional independence optimization yields
very significant gains in the performance of 
epistemic model checking. Depending on the 
example, the optimization yields a speedup 
as large as four orders of magnitude. 
Indeed, it can yield linear growth rates in  computation time on examples
that otherwise display an exponential growth rate. 
It adds significantly to the scale of the examples
that can be analyzed in reasonable time, 
increasing both the number of agents that can be 
handled, the length of their protocols, and 
the size of messages they communicate. 

The structure of the paper is as follows. 
Section~\ref{sec:background} provides background on 
the multi-agent epistemic logic that we consider, 
and on the epistemic model checking problem. 
An example of an application  of epistemic model checking, 
Chaum's Dining Cryptographers protocol \cite{chaum} is 
described in Section~\ref{sec:exampledc}.  
Section~\ref{sec:valalg} recalls Shenoy and Shafer's  
valuation algebra, which provides a general framework
for algorithms from both the database and Bayesian reasoning literature 
based on \emph{variable elimination}. A particular instance of this framework is introduced 
that is relevant to the present paper. Section~\ref{sec:condind-dg}
describes the notion of conditional independence
(for a discrete rather than probabilistic setting) that we use, 
and recalls ideas from the literature
that show how conditional independencies can be deduced
in models equipped with a directed acyclic graph structure. 
These ideas are then applied 
to our setting of epistemic model checking. 
Section~\ref{sec:dcopt} illustrates the application of these techniques
on the Dining Cryptographers problem. We then turn 
to describing our implementation of the optimization in MCK. 
Section~\ref{sec:symbeval} describes the symbolic evaluation 
method that relates program-based model checking inputs
to directed graphs and the overall structure of the 
implementation. Section~\ref{sec:results} gives the 
results of experiments that compare  performance of the 
optimized implementation of model checking with previous 
implementations. Section~\ref{sec:concl} concludes with a 
discussion of related work and future directions. 
Appendix A provides additional detail on the experiments. 

\section{Background: Epistemic Logic} 
\label{sec:background}

We begin by recalling some basic definitions from epistemic logic and 
epistemic model checking. We first define epistemic Kripke structures 
and a particular representation of them that we use in this paper, 
and then show how, in the context of model checking, an epistemic 
Kripke structure provides semantics for a multi-agent setting in 
which each agent's behaviour is described by a program. 

\subsection{Epistemic Kripke Structures} 

Let $\Prop$ be a set of atomic propositions, which we also call {\em variables}. 
An assignment for a set of variables $V$ is a mapping $\alpha: V\rightarrow \{0,1\}$. 
We write $\assgt(V)$ for the set of all assignments to variables $V$.
We denote the restriction of a function $f:S\rightarrow T$ to a subset $R$ of the domain $S$ 
by  $f\restrict R$. 

The syntax of epistemic logic for a set $\agts$ of agents is given by the grammar 
$$ \phi ::= p ~|~ \neg \phi ~|~ \phi \land \phi~|~ K_i\phi $$ 
where $p\in \Prop$ and $i\in \agts$.  That is, the language is 
a modal propositional logic with a set of modalities $K_i$, 
such that $K_i\phi$ means, intuitively, that the agent $i$ knows that $\phi$.   
We freely use common abbreviations from propositional logic, 
e.g., we write $\phi_1 \lor \phi_2$ for $\neg (\neg \phi_1 \land \neg \phi_2)$
and $\phi_1 \rimp \phi_2$ for $\neg \phi_1 \lor \phi_2$ and 
$\phi_1 \dimp \phi_2$ for $(\phi_1 \rimp \phi_2) \land (\phi_2 \rimp \phi_1)$. 
We write $\vars(\phi)$ for the set of variables occurring in the formula $\phi$.

Abstractly, an {\em epistemic Kripke structure} for a set of variables $\Prop$ is a 
tuple $\ks= (W,\sim, \pi)$ where $W$ is a set, $\sim= \{\sim_i\}_{i \in \agts}$ is a collection of equivalence relations $\sim_i$ on $W$, 
one for each agent $i$, 
and $\pi: W \rightarrow \assgt(V)$ is a function. 
Intuitively, $W$ is a set of possible worlds. The relation $u \sim_i v$ 
holds for $u,v\in W$ just when 
agent $i$ is unable to distinguish the possible worlds $u$ and $v$, 
i.e., when it is in the world $u$, the agent considers it to be possible
that it is in world $v$, and vice versa. 
For a proposition $p$, the value $\pi(u)(p)=1$  just when 
$p$ is true at the world $u$. We say that $\ks$ is \emph{finite} when it has a finite set of worlds.  

The semantics of epistemic logic is given by a ternary relation $\ks,w \models \phi$, 
where $\ks= (W,\sim, \pi)$ is a Kripke structure, $w\in W$ is a world of $\ks$, 
and $\phi$ is a formula. The definition is given recursively, 
by 
\be
\item $\ks,w \models p$ if $\pi(w)(p) = 1$, for $p \in \Prop$,  
\item $\ks,w \models \neg \phi$ if not $\ks,w \models \phi$,
\item $\ks,w \models \phi_1\land \phi_2$ if $\ks,w \models \phi_1$ and $\ks,w \models \phi_2$, 
\item $\ks,w \models K_i \phi$ if  $\ks,u \models \phi$ for all worlds $u\in W$ with $w\sim_i u$. 
\ee  
Intuitively, the clause for the operator $K_i$ says that $K_i\phi$ holds when $\phi$ is true at
all worlds that the agent considers to be possible.  
We write $\ks \models \phi$ when $\ks,w\models \phi$ for all worlds $w\in W$.

For two Kripke structures $\ks = (W,\sim,\pi)$ and $\ks' = (W',\sim',\pi')$, 
a \emph{bisimulation} with respect to a set of variables $U\subseteq \Prop$ 
is a binary relation $\bisim \subseteq W \times W'$
such that:
\be 
\item (atomic) If $u \bisim u'$ then $\pi(u) \restrict U= \pi(u') \restrict U$. 
\item (forth) If $u \bisim u'$ and $u \sim_i v$ then there exists $v'\in W'$ such that $u' \sim_i v'$ and $u' \bisim v'$. 
\item (back) If $u \bisim u'$ and $u' \sim_i v'$ then there exists $v\in W$ such that $u \sim_i v$ and $v \bisim v'$. 
\ee 
If there exists a bisimulation whose projection on the first component is $W$, and on the second component is $W'$, 
then we say that the structures are \emph{bisimilar} with respect to $U$, and write $\ks \bisim_U \ks'$. 
The following result is well-known in modal logic \cite{vanBenthem}.

\begin{proposition}  \label{prop:bisim}
If $\bisim$ is a bisimulation with respect to $U$ and $w \in W, w'\in W'$ are worlds with $w\bisim w'$, 
then for all formulas $\phi$ over atomic propositions $U$, we have $\ks,w \models \phi$ iff $\ks',w'\models \phi$. 
Moreover, if  $\ks \bisim_U \ks'$ then $\ks \models \phi$ iff $\ks'\models \phi$. 
\end{proposition} 

It will be convenient to work with a more concrete representation of 
Kripke structures that treats worlds as assignments to variables. 
For simplicity, we assume that all variables are boolean. 

Define an {\em epistemic variable structure} over a set of variables $V $ to be a 
tuple $\vs = (A, O,V)$ where   $A \subseteq \assgt(V)$ and 
$O = \{O_i\}_{i\in \agts}$ is a collection of sets of variables 
$O_i\subseteq V$, one for each agent $i$.
Intuitively, such a structure is an alternate representation of a Kripke structure, 
where the indistinguishability relation for an agent is specified by means of a set of variables
observable to the agent. 

Given an epistemic variable structure $\vs = (A,O,V)$, 
we obtain a Kripke structure $\vstoks(\vs) = (W,\sim, \pi)$, 
with $W = A$. The relation $\sim_i$ for agent $i$ is defined by 
$u \sim_i v$ when $ u \restrict O_i = v\restrict O_i$.  
The assignment $\pi$ is defined by $\pi(w) = w$. 

Conversely, any (finite) Kripke structure $\ks = (W,\sim,\pi)$ over variables $\Prop$
can be represented as an epistemic variable structure that satisfies the 
same set of formulas over  $\Prop$, but may use a larger set of variables to represent states. 
The construction uses two sets of additional variables.%
\footnote{We give a simple construction here, but note that the result can be proved using a 
smaller set of additional variables, by encoding equivalence classes in binary.}
 For 
each equivalence class $c=[w]_i$ of the equivalence relation $\sim_i$, 
we define a proposition $p_{c,i}$ whose meaning, intuitively, is that the 
current world is in the class $c$ of $\sim_i$. Let $V_\sim = \{ p_{[w]_i,i}~|~ w \in W,~i \in \agts\}$. 
For each world $w$, we also define a proposition $p_w$ that means, intuitively, 
that the  current world is $w$. Let $V_W =  \{ p_{w}~|~ w \in W\}$. 
Let $V_M = V \cup V_\sim\cup V_W$. 

We extend the assignment $\pi(w)$, which has domain $V$, 
to an assignment $\pi_w$ with domain $V_M$, by defining 
\bi
\item $ \pi_w(p) =  \pi(w)(p)$, for $p\in V$,  and
\item $\pi_w(p) =1 $ iff  $w \in [u]_i$, for  $p= p_{[u]_i,i} \in V_\sim$, and 
\item $\pi_w(p) =1 $ iff $w = u$, for $p= p_{u} \in V_W$. 
\ei 
Write $A_W$ for the set $\{\pi_w~|~ w\in W\}$. 
Define $\kstovs(\ks)$ to be the epistemic variable structure $(A_W, O)$, 
where $O_i = \{ p_{[w]_i,i}~|~w\in W\}$ for each $i \in \agts$.  
That is, the assignments in this structure are the 
extended assignments $\pi_w$, and we take the set of observable variables to be 
precisely the set of variables $V_\sim$ representing equivalence classes. 

\begin{proposition} If $\ks$ is a Kripke structure over variables $V$, then 
$\ks$ is bisimilar to $\vstoks(\kstovs(\ks))$ with respect to $V$.
\end{proposition} 

\begin{full} 
\begin{proof} 
When $\ks $ has worlds $W$, the Kripke structure $\vstoks(\kstovs(\ks))$
has the same set of worlds as $\kstovs(\ks)$, i.e., $A_W$. 
Consider the relation $\bisim \subseteq W \times A_W$ defined
by $\bisim = \{ (w,\pi_w)~|~w\in W\}$. We show that this is a bisimulation
between $\ks$ and $\vstoks(\kstovs(\ks))$. 

Note first that $\bisim$ sets up a 1-1 correspondence 
between $W$ and $A_W$, since  if $u,v\in W$ with $u \neq v$ then 
$\pi_u(p_v) =0$ and $\pi_v(p_v) = 1$, so $\pi_u \neq \pi_v$. 
Thus, $w \bisim \pi_u$  implies $u = w$, so $\pi(w)(p) = \pi_u(p)$ for all $p\in V$. 
This gives condition (atomic). 

We show that for $u,v\in W$, and $i \in \agts$, we have $u \sim_i v$ iff $\pi_u \sim_i \pi_v$
(i.e., for all $p \in O_i$, we have $\pi_u(p) = \pi_v(p)$). 
In particular, note $u \sim_i v$ implies $[u]_i = [v]_i$, 
so for all $p = p_{[w]_i,i}  \in O_i$, we have $\pi_u(p)=1$ iff 
$u \in [w]_i$ iff $u \sim_i w$ iff $v\in [w]_i$ iff $\pi_v(p) =1$. 
Conversely, if  for all $p \in O_i$, we have $\pi_u(p) = \pi_v(p)$, 
then $\pi_u(p_{[v]_i,i}) =  \pi_v(p_{[v]_i,i}) =  1$, since $v \in [v]_i$, 
and it follows by definition that $u \in [v]_i$, i.e., $u\sim_i v$. 

The conditions (forth) and (back) now follow straightforwardly. 
For (forth), note that if $u\sim v$ and $u\bisim u'$, 
then $u' = \pi_u$. Taking $v' = \pi_v$, we have $v \bisim v'$,  
and $u' =\pi_u \sim \pi_v = v'$  from the above. 
The proof of (back) is similar.  
\end{proof}
\end{full}  

Using Proposition~\ref{prop:bisim}, it follows that for all formulas $\phi$, we have 
$\ks \models \phi$ iff $\vstoks(\kstovs(\ks)) \models \phi$. 
Thus, for purposes of the modal language, it suffices to work with epistemic variable 
structures in place of finite Kripke structures.  Henceforth, for an epistemic variable structure 
$\vs$, and world $w$ of $\vs$, we write 
$\vs, w \models \phi$ if $\vstoks(\vs), w \models \phi$ and $\vs \models \phi$ if $\vstoks(\vs) \models \phi$.

\newcommand{\atl}{\langle} 
\newcommand{\atr}{\rangle} 
\newcommand{\skp}{\mathit{skip}}

\subsection{From Programs to Epistemic Kripke Structures} 

In the context of model checking, one is interested in analyzing a model 
represented as a program. We now show how programs generate a Kripke structure that serves as their semantics. 
We work with a very simple straightline programming language in which a multi-agent scenario is 
represented by each of the agents running a protocol in the context of an environment. 
The syntax and operational semantics  of this language  is shown in Figure~\ref{fig:lang}. 

Intuitively, all variables (represented by non-terminal $v$) in this fragment are boolean, and $e$ represents a boolean expression. 
Code $C$ consists of a sequence of assignments and randomization statement $rand(v)$, which assigns a random value to $v$. 
Non-terminal $a$ 
represents an atomic action, either the skip statement $\skp$, or an atomic statement 
$\atl C\atr$  consisting of code $C$ that executes without interference
from code of other agents.  An agent protocol $P$ consists of a sequence 
of atomic actions: protocol $\epsilon$ represents termination, and 
is treated as equivalent to $\skp;\epsilon$ to capture that a terminated
agent does nothing while other agents are still running. 
A \emph{joint protocol} $J$, is 
represented by a statement of the form $P_1 ~||~ \ldots ~||~P_n~\Delta~C_E$, 
and consists of a number of agent protocols $P_1,\ldots,P_n$, 
running in the context of an environment represented by code $C_E$. 

There are two relations in the operational semantics. 
States $s$ are assignments of boolean variables to boolean values, 
and we write $e(s)$ for the value of boolean expression expression $e$ 
in state $s$. The binary relation $\rightarrow_0$ on configurations of type $(s,C)$
represents \emph{zero-time} state transitions, which do not change
the system clock. The binary relation $\rightarrow_1$ 
on configurations of type $(s,J)$  represents state transitions 
corresponding to a single clock tick. 
Thus, 
$C\rightarrow_0^* \epsilon$ represents that code $C$  runs to termination in time $0$. 
In a single tick transition represented by $\rightarrow_1$, we take the next 
atomic action $a_i = \atl C_i\atr$ from each of the agents, 
and compose the code $C_i$ in these actions  with the code from the environment 
$C_E$ to form the code $C= C_1;\ldots C_n; C_E$. The single step transition
is obtained as the result of running this code $C$ to termination in
zero-time. 

\begin{figure} 
$$\begin{array}{l} 
e ::= v ~|~ \neg v~| ~v\land v ~|~ v \lor v~| ~ \ldots \\ 
C::= \epsilon ~|~ v := e ; C~|~ rand(v) ; C \\ 
a ::= \atl C\atr ~|~ \skp \\ 
P ::= \epsilon ~|~   a ; P\\ 
J ::= P~ ||~ ... ~||~ P~\Delta ~ C 
\end{array} 
$$

$$\begin{array}{ccc}
 (s, v:=e;C) \rightarrow_0 (s[e(s)/v], C) & \quad \quad & 
(s, \skp;C) \rightarrow_0 (s,C) \\[10pt]
(s, rand(v);C) \rightarrow_0 (s[0/v], C) & & 
(s, rand(v);C) \rightarrow_0 (s[1/v] ,C) 
\end{array} \\[10pt] 
$$

$$\begin{array}{c} 
a_1 = \atl C_1 \atr  ~ \ldots  ~a_n = \atl C_n \atr ~~~~C= C_1 ; \ldots ; C_n; C_E~~~~(s, C) \rightarrow_0^* (t, \epsilon)\\  
\hline 
(s,  ~a_1 ;P_1 ~||~ \ldots  ~||~ a_n;P_n ~ \Delta~ C_E) \rightarrow_1  (t, ~P_1 ~|| ~\ldots ~||~ P_n  ~\Delta~ C_E )
\end{array} 
$$
\caption{Syntax and Operational Semantics of Programs\label{fig:lang}} 
\end{figure} 

A {\em system}  is represented using this programming language by means of a
tuple $\sys = (J,I,O)$, where $J$ is a joint protocol for $n$ agents, 
$I$ is a boolean formula expressing the initial condition, 
and $Q$ is a tuple of $n$ sets of variables, with $Q_i$ representing
the variables observable to agent $i$. 

Given a maximum running time $n$, a system $\sys=(J,I,Q)$ is associated to an epistemic variable structure 
$\vs_n(\sys)=\langle A,O, V\rangle$ as follows. A 
\emph{run} of length $n$ of the system is a sequence of states $r= s_0, s_2, \ldots, s_n$, where 
$s_0$ satisfies the initial condition $I$ and 
$(s_0,J) \rightarrow_1 (s_1,J_1) \rightarrow_1 \ldots \rightarrow_1 (s_n,J_n) $ for some $J_1, \ldots, J_n$.  
If $U$ is the set of variables appearing in $J$, 
we define $V$ to be the set of \emph{timed variables}, 
i.e., the set of variables $v^t$ where $0\leq t \leq n$.  
We take  $A$  to be the set of assignments $\alpha_r$ to variables $V$ 
derived from runs $r$ by $\alpha_r(v^t) = s_t(v)$ when $v\in U$ and $0\leq t \leq n$. 
For the perfect recall semantics, which is our focus in this paper, 
we define the observable variables $O_i$ for agent $i$ 
to be the set of timed variables $v^t$ where $v\in Q_i$ and $0 \leq t\leq n$.

\section{Example: Dining Cryptographers} 
\label{sec:exampledc} 
We illustrate epistemic model checking and the optimizations developed in this paper 
using Chaum's Dining Cryptographers Protocol \cite{chaum},  a security protocol whose aim is 
to achieve an anonymous broadcast. This protocol, both in its basic form, as well as an extension that is
more generally applicable, has previously been analysed
using epistemic model checking \cite{MS,AlBatainehMeyden10}. Chaum introduces the protocol  with the following story: 

\begin{quote}
Three cryptographers are sitting down to dinner at their favourite restaurant.
Their waiter informs them that arrangements have been made with the maitre d'hotel
for the bill to be paid anonymously. One of the cryptographers might be paying for
the dinner, or it might have been NSA (U.S. National Security Agency). The three
cryptographers respect each other's right to make an anonymous payment, but they
wonder if NSA is paying. They resolve their uncertainty fairly by carrying out the following protocol:\\

Each cryptographer flips an unbiased coin behind his menu, between him and the cryptographer on his right, so that only the two of them can see the outcome. Each cryptographer then states aloud whether the two coins he can see--the one he flipped and the one his left-hand neighbor flipped--fell on the same side or on different sides. If one of the cryptographers is the payer, he states the opposite of what he sees. An odd number of differences uttered at the table indicates that a cryptographer is paying; an even number indicates that NSA is paying (assuming that the dinner was paid for only once). Yet if a cryptographer is paying, neither of the other two learns anything from the utterances about which cryptographer it is.
\end{quote}

\newcommand{\paid}{\mathit{paid}} 
\newcommand{\coin}{\mathit{coin}}
\newcommand{\leftcoin}{\mathit{left}}
\newcommand{\say}{\mathit{say}}
\newcommand{\pinit}{\mathit{pinit}}
  
The solution generalizes to any number $n$ of cryptographers $C_0, \ldots , C_{n-1}$ 
at the table.  We may represent the protocol by means of the following program for 
cryptographer $i$, who is assumed to have a boolean variable $\paid_i$ 
that indicates whether (s)he is the payer. (The program starts running from an initial state
in which the  constraint $\bigvee_{0\leq i < j \leq  n-1} \neg( \paid_i\land \paid_j)$  is satisfied.) 
We write $\oplus$ for the exclusive-or. 

\begin{tabbing} 
nnn \= \kill 
$C_i$: \\
Observed variables: $\paid_i$, $\coin_i$, $\leftcoin_i$, $\say_0, \ldots , \say_{n-1}$\\ 
Protocol: \\ 
\> $\coin_i := rand$ ; \\ 
\> $\leftcoin_{i+1~mod~n} := \coin_i$ ; \\
\> $\say_i := \paid_i \oplus \coin_i \oplus \leftcoin_i$ 
\end{tabbing} 

All variables take boolean values. Here $rand$ is the generation of a random
boolean value: in a probabilistic interpretation, the value would be drawn from a uniform 
distribution, but for our purposes in epistemic model checking, we
interpret this operation as nondeterministically selecting a value of either $0$ or $1$. 
Each cryptographer is associated with a set of variables, whose values they are  
able to observe at each moment of time. Note that a cryptographer may write to 
a variable that they are not able to observe. In particular, $C_i$ writes to the variable 
$\leftcoin_{i+1~mod~n}$ that is observed only by $C_{i+1~mod~n}$.

\begin{figure} 
\centerline{\includegraphics[height=10cm]{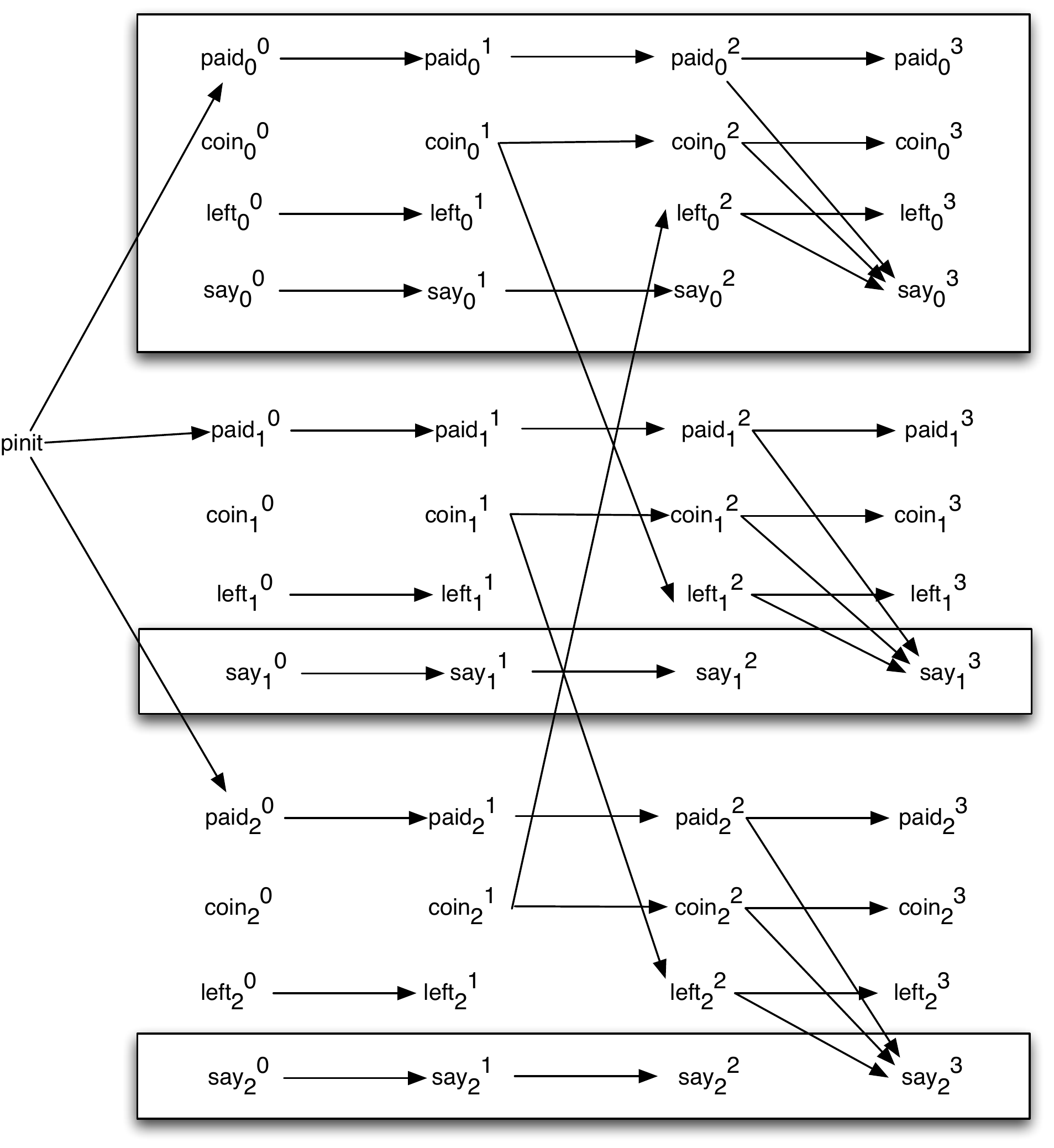}}
\caption{Timed-variable dependency graph after program unfolding \label{fig:initgraph}} 
\end{figure}

We will work with {\em dependency networks} that  show how the values of 
variables change over time. 
The  DC protocol runs for 4 ticks of the clock, 
(time 0 plus one tick for each step in the protocol), so we have instances $v^0 \ldots v^3$ 
of each variable $v$. Figure~\ref{fig:initgraph} shows the dependencies between 
these instances. The figure is to be understood as follows: a variable $v^t$ takes a value
that directly depends on the values of the variables $u^{t-1}_1 \ldots u^{t-1}_n$ such that there is 
an edge from $u^{t-1}_j$ to $v^t$.  Additionally, there is a dependency between the initial 
values $\paid_i^0$ captured using a special variable $p_{init}$. (We give a more formal presentation of such dependency structures  below.)  
The observable variables for agent $C_0$ have been indicated by rectangles: timed variables inside
these rectangles are observable to $C_0$. 

\section{Valuation Algebra} \label{sec:valalg} 

Shenoy and Shafer \cite{Shenoy89,ShenoyShafer90}  have developed a general axiomatic formalism that captures the 
key properties that underpin the correctness of optimization methods used for a variety of 
uncertainty formalisms. In particular, it has been shown that this 
formalism allows for a general explanation of variable elimination 
algorithms and the notion of conditional independence used in 
the Bayesian Network literature \cite{KollerFriedman}, and applies
also in other contexts such as Spohn's theory of \emph{ordinal conditional 
functions} \cite{Spohn88}.  There is a close connection also to ideas 
in database query optimization \cite{Maier83} and operations research \cite{BB72}. 
We show here that Shenoy and Shafer's general axiomatic framework 
applies to 
epistemic model checking. 
This will enable us to apply the variable elimination algorithm  
to derive techniques for optimizing epistemic model checking. 

\newcommand{\Vars}{\mathit{Vars}} 
\newcommand{\Valns}{\Phi} 
\newcommand{\dom}[1]{\mathit{dom}(#1)} 
\newcommand{\val}[1]{\Omega_#1}

\newcommand{\comb}{\otimes} 
\newcommand{\marg}{\downarrow} 
\newcommand{\Doms}{D} 
\newcommand{\valo}{s}
\newcommand{\valt}{t}
\newcommand{\idty}{e}

\subsection{Axiomatic Framework} 
\label{sec:valalg_ax}

We begin by presenting Shenoy and Shafer's framework, following \cite{KohlasShenoy}. 
Let $\Vars$ be a set of variables, with each $v \in \Vars$ taking values in a set $\val v$. 
For a  set $X$ of variables, the set  $\val{X} = \Pi_{x\in X} \val{x}$
is called the \emph{frame} of $X$. Elements of $\val{X}$  are called 
\emph{configurations} of $X$. In case $X=\emptyset$, the set 
$\val{X}$ is interpreted as $\{\langle\rangle\}$, i.e., the set 
containing just the empty tuple. We write $\Doms$ for ${\cal P}(\Vars)$.

A \emph{valuation algebra} is a tuple $\langle \Phi, \mathit{dom}, e, \comb,\marg\rangle$, 
with components as follows.  A state of information is represented in valuation algebra 
by a primitive notion called a \emph{valuation}. Component $\Phi$ is a set, the set of all valuations, 
and $\mathit{dom}$ is function from $\Phi$ to ${\cal P}(\Vars)$. 
Intutitively, for each valuation $s\in \Phi$, the domain $\dom{s}$ is the set of 
variables that the information is about. For a set of variables $X$, we write $\Valns_X$
for the set of valuations  $s$ with $\dom{s} =X$.  
Component $\idty$ gives an element $\idty_X\in \Phi_X$ for each $X\in D$. 
A valuation algebra also has two operations $\comb: \Valns \times \Valns \rightarrow \Valns$ (combination) and $\marg: \Valns\times \Doms \rightarrow \Valns$(marginalization), 
with  $\comb$, intuitively, representing the combination of two pieces of 
information, and $\marg$   used to restrict a piece of information to a given set
of variables.  Both are written as infix operators. From marginalization, another operator $-: \Valns\times \Vars \rightarrow \Valns$ 
called {\em variable elimination} can be defined, by $\valo^{-x} = \valo \marg (\dom{\valo} \setminus \{x\}) $. 

These operations are required to satisfy the following conditions: 
\be
\item[VA1.]  {\em Semigroup.} $\comb$ is associative and commutative. 
For all $X\in  \Doms$ and  all
$\valo\in \Valns_X$, we have $\valo\comb \idty_X = \idty_X\comb \valo = \valo$. 

\item[VA2.] {\em Domain of combination.} For all $\valo,\valt\in\Valns$, $\dom{\valo \comb\valt} = \dom{\valo} \cup \dom{\valt}$. 

\item[VA3.]{\em Marginalization.} For $\valo \in \Valns$ and $X,Y\in \Doms$, the following hold:  
$$ 
\valo\marg X = \valo \marg X\cap \dom{\valo} \quad\quad 
\dom{\valo \marg X} = X\cap \dom{\valo} \quad\quad 
\valo \marg \dom{\valo} = \valo ~.
$$ 

\item[VA4.] {\em Transitivity of marginalization.} For $\valo \in \Valns $, and $X\subseteq Y\subseteq \Vars$, 
$$ (\valo \marg Y)\marg X = \valo\marg X ~.$$ 

\item[VA5.]  {\em Distributivity of marginalization over combination.} 
For $\valo,\valt\in \Valns$, with $ \dom{\valo} = X$ 
$$ (\valo \comb \valt) \marg X = \valo \comb (\valt\marg X)~.$$

\item[VA6.] {\em Neutrality.} For $X, Y\in \Doms$, 
$$ \idty_X \comb \idty_Y = \idty_{X\cup Y}~. $$  

\ee 

\newcommand{\Fus}{\mathit{Fus}}
\newcommand{\bigcomb}{\bigotimes}

A key result that follows from these axioms, called the 
\emph{Fusion Algorithm} \cite{Shenoy92}, exploits 
Distributivity of Elimination over Combination to give 
a way of computing the result of a marginalization operation 
applied to a sequence of combinations, by 
\emph{pushing in} variable eliminations over elements of the 
combination that do not contain the variable. 

For a finite set $S = \{\valo_1, \valo_2, \ldots, \valo_k\} \subseteq \Valns$, 
write $\comb S$ for $\valo_1 \comb \valo_2 \comb \ldots \comb \valo_k$. 
We define the \emph{fusion} of $S$ via $x\in \Vars$ to be 
the set $$\Fus_x(S) = \{ (\comb S_+)^{-x}\} \cup S_-~$$
where we have partitioned $S$ as $S_+\cup S_-$, such that 
$S_+$ is the set of $\valo\in S$ with $x\in \dom{\valo}$, 
and $S_-$ is the set of $\valo\in S$ with $x\not \in \dom{\valo}$. 
That is, in the fusion of the set $S$ with respect to $x$, we combine all the
valuations with $x$ in their domain, and then eliminate $x$, and
preserve all valuations with $x$ not in their domain. 

Suppose we are interested in computing 
$(\comb S)\marg X$, for $S$ a finite set of valuations, 
and $X\subseteq \Vars$. The Fusion Algorithm 
achieves this by repeatedly applying the fusion 
operation, using some ordering of the variables in $X$. 
We write $\dom{S}$ for $\dom{\comb S} = \bigcup \{ \dom{\valo}~|~ \valo \in S\}$

\begin{theorem}[\cite{Shenoy92}] 
Let $S$ be a finite set of valuations, and $X\subseteq \Vars$. 
Suppose  $\dom{S} \setminus X = \{x_1,x_2, \ldots , x_n\}$. 
Then 
$$ (\comb S) \marg X = \comb \Fus_{x_n}( \ldots (\Fus_{x_1}(S))) ~.$$  
\end{theorem} 

Each ordering of the variables $x_1 \ldots x_n$ gives 
a different way to compute $(\comb S) \marg X$. 
A well chosen order can yield a significant optimization 
of the computation, by keeping the domains of the intermediate valuations in the sequence of fusions small. 
Finding an optimal order may be  computationally complex, but there exist heuristics that 
produce good orders in practice \cite{Olmsted,Kong}.

\subsection{A Valuation Algebra of Relational Structures} 
\label{sec:valrel} 

We now show that the relational structures that underly Kripke structures are associated  with algebraic 
operations that satisfy the conditions from the previous section. It will follow from this 
that the Fusion algorithm can be applied to these structures. 

\newcommand{\AllProp}{{\cal V}}
\newcommand{\rels}{s}

Let $\AllProp$ be the set of all variables. 
Values in the algebra will be \emph{relational structures} of the form $\rels = (A,V)$, 
where $V\subseteq \AllProp$ and $A\subseteq \assgt(V)$. 
The domain of a relational structure is defined to be its set of variables, 
i.e. if $\rels = (A,V)$ then $\dom{\rels} = V$. 
We define the identities $\idty_X$ and operations $\comb$  of combination and and $\marg$ of marginalization as follows. 
Let $\rels_1 = (A_1, V_1)$ and $\rels_2 = (A_2, V_2)$ and $X \subseteq \AllProp$. 
Then 
\begin{itemize} 

\item $\idty_X = (\assgt(X),X)$,

\item $\rels_1 \comb \rels_2 = (A,V)$
where $V= V_1\cup V_2$,
and $A\subseteq \assgt(V)$ is defined by $\alpha \in A$ iff $\alpha\restrict V_1 \in A_1$ and $\alpha\restrict V_2 \in A_2$. 
\item $\rels_1 \marg X = (A,V)$ where 
$V = V_1 \cap X$, 
and $A = \{ \alpha\restrict X~|~ \alpha\in A_1\}$.
\end{itemize} 
To use terminology from relational databases, $\rels_1\comb \rels_2 $ is the join of
relations and $\rels \marg X $ is the projection of the relation $\rels$ onto attributes $X$.  
The following result is straightforward; these properties are well-known for 
relational algebra. 

\begin{proposition} 
The algebra of relational structures satisfies axioms VA1-VA6. 
\end{proposition}   

We may extend the operation of marginalization in this valuation algebra
to epistemic variable structures as follows. If $\vs= (A,O,V)$
is an epistemic variable structure and $X\subseteq V$, we define $\vs\marg X = (A',O',V')$
where $A' = \{\alpha\restrict X~|~ \alpha \in A\}$
and $O'_i = O_i \cap X$ for all $i\in \agts$ and $V' = V\cap X$. 
In general, this operation results in agents losing information, since their
knowledge is based on the observation of fewer variables. Below, we
identify conditions where knowledge is preserved by this operation.

\section{Conditional Independence and Directed Graphs}
 \label{sec:condind-dg}

\subsection{Conditional Independence} 
 \label{sec:condind}

Let $X,Y,Z\subseteq \Prop$ be sets of variables. The notion of conditional 
independence expresses a generalized type of independency relation. 
Variables $X$ are said to be conditionally independent of $Y$, given $Z$, 
if, intuitively, once the values of $Z$ are known, the values of $Y$ are 
unrelated to the values of $X$, so that neither $X$ not $Y$ gives
any information about the other. This intuition can be formalized for both 
both probabilistic and discrete models. The following definition gives 
a discrete interpretation, related to the notion of 
\emph{embedded multivalued dependencies} from database theory \cite{Fagin77}.

\begin{definition} 
Let $A\subseteq \assgt(V)$ be a set of assignments  over variables $\Prop$ and 
let $X,Y,Z \subseteq \Prop$. We say that $A$ satisfies the conditional independency
$X \bot Y|Z$, and write $A \models X\bot Y|Z$, if 
for every pair of worlds $u,v \in A$ with $u\restrict Z = v\restrict Z$, there exists 
$w \in A$ with $w\restrict X\cup Z = u \restrict X\cup Z$ and $w\restrict Y\cup Z  = v\restrict Y\cup Z$. 
For an epistemic variable structure $\vs= (A,O,V)$, we write $\vs  \models X\bot Y|Z$ if 
$A \models X\bot Y|Z$. 
\end{definition} 

Conditional independencies can be deduced from graphical representations of models. 
Such representations have been used in the literature on Bayesian 
Nets \cite{pearlprobbook,KollerFriedman}, and have also been applied in propositional reasoning \cite{Darwiche97,Darwiche98}. 
The following presentation is similar to \cite{Darwiche97} except that we work with relations over arbitrary domains rather than 
propositional formulas. 

\newcommand{\arr}{\rightarrow}
\newcommand{\und}{-}
\newcommand{\parents}[1]{\mathit{pa}(#1)}
\newcommand{\ancestors}[1]{\mathit{An}(#1)}

\subsection{Directed Graphs}

A \emph{directed graph} is a tuple $G=(V,E)$ consisting of a set $V$ (the vertices) 
and a relation $E\subseteq V\times V$ (the edges). If $(u,v)\in E$ we say that there is 
an edge from $u$ to $v$, and may also denote this fact by $u \arr v$. 
We write $u \und v$ when both $u\arr v$ and $v \arr u$. 
The set of \emph{parents} of a node $u$ is defined to be the set $\parents v = \{u\in V~|~ u\arr v\}$.  
A \emph{path of length $n$ from $u$ to $v$} in $G$ is a sequence $u_0, u_1, \ldots, u_n$  of vertices, 
such that $u_i \arr u_{i+1}$ for all $i= 0\ldots n-1$. The graph is \emph{acyclic} if there is no nontrivial 
path from any vertex to itself. We also call such a graph a directed acyclic graph, abbreviated as \emph{dag}. 
An \emph{undirected graph} is a graph with a symmetric edge relation $E$, i.e. if $u\arr v$ then also $v \arr u$. 
We may represent such a pair of edges with the notation $u \und v$.

The notion of d-separation \cite{pearlprobbook} provides a way to derive a set of 
independency statements from a directed graph $G$. We present here an equivalent formulation
from \cite{LDLL90}, that uses the notion of the \emph{moralized} graph $G^m$ of a directed graph $G$. 
The graph $G^m$ is defined to be the undirected graph obtained from $G$ by first adding an edge 
$u \und v$ for each pair $u,v$ of vertices that have a common child (i.e. such that 
there exists $w$ with $u\arr w$ and $v \arr w$), and then replacing all directed edges with undirected
edges.  For a set of vertices $X$ of the directed graph $G$, we write $\ancestors{X}$ for the set of 
all vertices $v$ that are ancestors of some vertex $x$ in $S$ (i.e., such that there 
exists a directed path from $v$ to $x$). 
For a subset $X$ of the set of vertices of graph $G=(V,E)$, we defined the restriction of
$G$ to $S$ to be the graph $G_S = (V\cap S,\{(u,v) \in E~|~ u,v\in S\})$.  
For disjoint sets $X,Y,S$, we then 
have that \emph{$X$ is d-separated from $Y$ by $S$} if all paths from $X$ 
to $Y$ in $(G_{\ancestors{X\cup Y\cup S}})^m$ include a vertex in $S$.

A \emph{structured model} for a valuation algebra $\langle \Phi, \mathit{dom}, e, \comb,\marg\rangle$,
is a tuple $M= \langle V,E, \relss \rangle$   
where $V$ is a set of variables, $D = \pow{V}$, component $E$ is a binary relation on $V$ such that $G_M = (V,E)$ is a dag,  and $\relss= \{\rels_v\}_{v\in V}$
 is a collection of values in $\Phi$ such that for each variable $v\in V$, we have 
 \begin{itemize} 
 \item 
 $\dom{\rels_v} = \{v\} \cup \parents v$, i.e. the domain of $\rels_v$  
 consists of $v$ and its parents in the dag, 
\item 
$\rels_v \marg {\parents v }= \idty_{\parents{v}}$.  
\end{itemize} 
Intuitively, the second constraint says that the relation $\rels_v$ does not constrain the parents of $v$: for each 
assignment of values to the parents of $v$, there is at least one value
of $v$ that is consistent. 

The following is a consequence of results in \cite{pearlprobbook,LDLL90,VP88}. 

\begin{proposition}
Suppose that $M= \langle V,E, \relss\rangle$   is a structured 
model 
and $X,Y,Z$ are disjoint subsets of the vertices $V$ of the directed graph $G = (V,E)$. 
If $X$ is d-separated from $Y$ by $Z$, then $\comb \relss \models X\bot Y|Z$. 
\end{proposition}

\begin{full} 
\begin{proof} (Sketch)
The set $I$ of conditional independency statements holding in a structured model is a 
\emph{semi-graphoid}. 
In particular, we will take $I$ to be the semi-graphoid of conditional independencies in $\comb \relss$. 

A \emph{stratified protocol $L$ of $I$} is an ordering $v_1 \ldots v_n$ of $V$  
together with a function $p: V \rightarrow {\cal P}(V)$ such that for 
all $i = 1 \ldots n$, we have that $p(v_i) \subseteq \{ v_1, \ldots , v_{i-1}\}$ and 
the set $I$ contains the statement 
$$\{v_1 \ldots v_{i-1}\} \setminus p(v_i) \bot \{v_i\} | p(v_i)~.$$
Each stratified protocol $L$ is associated with a directed acyclic graph $dag(L) = (V,\{(u,v)~|~ v\in V, ~u \in p(v)\})$. 
It follows from the fact that $M$ is a structured relational model that any 
topological sort of $G$, together with the parent function in $G$ as the function $p$, 
is a stratified protocol $L$ of $I$, and we have $dag(L) = G$.  

Verma and Pearl \cite{VP88}(Theorem 2) show that if $Z$ d-separates $X$ from $Y$ in $dag(L)$ 
and $L$ is a stratified protocol for a semi-graphoid $I$, then 
$X\bot Y|Z$ is in $I$. It follows that  if $Z$ d-separates $X$ from $Y$ in $G$ then
$\comb\relss \models X\bot Y|Z$.  
\end{proof} 
\end{full} 

Structured models have an additional property that provides an 
optimization when eliminating variables: if a leaf node is
one of the variables eliminated from the combination of the nodes of the graph, 
then it can be simply removed from the model without changing the result. 
This is formally captured in the following result. 

\begin{proposition} 
Suppose that $M= \langle V,E, \relss\rangle$   is a structured  model, 
let $X\subseteq V$ and let $v \in V\setminus X$ be a leaf node. 
Then $\comb \relss \marg X = \comb (\relss \setminus \{s_v\})\marg X$.  
\end{proposition} 

\begin{full} 
\begin{proof}
Let $s = \comb( \relss \setminus \{s_v\})$.  By the semigroup properties, we have $\comb \relss = s \comb s_v$. 
We first note that 
\begin{align*} 
(s\comb s_v) \marg \dom{s} & = s\comb (s_v \marg \dom{s}) &  \text{by  VA5}\\
& = s\comb( s_v   \marg \dom{s} \cap \dom{s_v}) & \text{by VA3} \\ 
& = s\comb( s_v   \marg \parents{v}) & \text{since $v$ is a leaf} \\ 
& = s\comb e_v & \text{since $M$ is a s.r.m.} \\ 
& = s & \text{by VA1.} \\ 
\end{align*} 
Hence 
\begin{align*} 
(s\comb s_v) \marg X & = (s\comb s_v) \marg\dom{s}\cap X & \text{since $X\subseteq \dom{s}$}\\
& = (s\comb s_v) \marg \dom{s} \marg  X & \text{by VA4}\\
& = s  \marg  X & \text{by the result above.}\\
\end{align*} 
\end{proof}
\end{full}  

To apply these results for structured models to model checking epistemic logic, 
we use the following definition. We say that a structured model  $M= \langle V,E, \relss\rangle$ 
\emph{represents} the worlds of an epistemic variable structure $\vs=(A,O,U)$ 
if $V=U$ and $A = \comb \relss$. That is, the structured model 
captures the set of assignments making up the epistemic 
variable structure. 

\subsection{Eliminating  Observable Variables} \label{sec:elimovars}

Consider the following formulation of the model checking problem: for an epistemic formula $\phi$, we wish to 
verify  $\vs\models \phi$ where $\vs = (A,\Ovars,V)$ is an 
epistemic variable  structure  with observable variables $\Ovars$, with worlds 
represented by a  structured model $M= \langle V,E, \relss\rangle$. 

A first idea for how to optimize this verification problem is to reduce the structure 
$\vs$ to the set of variables $\vars(\phi)$  of the formula $\phi$, on the intuition 
that only these variables are relevant to the satisfaction of $\phi$. 
But this is not quite correct: the formula may contain the epistemic operators $K_i$, 
the semantics of which refers to the observable variables $\Ovars_i$, 
since these are used to define the indistinguishability relation. Thus a more 
accurate claim is that we should restrict the structure to 
$\vars(\phi)$, together with the sets $\Ovars_i$ for any operator $K_i$ in $\phi$. 

In fact, using the notion of  conditional dependence, 
it is often possible to identify a smaller set of  variables 
that suffices to verify the formula. The intuition for this 
is that some of the observed variables in $\Ovars_i$ may be independent of the 
variables in the formula, and moreover, information may be redundantly 
encoded in the observable variables. For example, if an observable variable
that does not itself occur in the formula
is computed from other observable variables, then it is redundant from the 
point of view of determining the possible values of variables in the 
formula.  The following definitions strengthen the idea of restricting to 
$\vars(\phi) \cup \Ovars$ by exploiting a sufficient condition for
the removal of observable variables.

\newcommand{\keep}{\kappa}

Say that $\keep$ is a \emph{relevance} function for a formula $\phi$ with respect to an
epistemic variable structure $\vs= (A, O, V)$ if it maps subformulas of $\phi$ to
subsets of the set of variables $V$, and satisfies the following  conditions: 
\be 
\item $\keep(p) = \{p\}$ for $p \in \Prop$, 
\item $\keep(\phi_1 \land  \phi_2 )=  \keep(\phi_1) \cup \keep(\phi_2)$, 
\item $\keep(\neg \phi_1) = \keep(\phi_1)$,  and 
\item $\keep(K_i\phi_1) = U_i \cup \keep(\phi_1)$, for some 
$U_i \subseteq O_i$ with $\keep(\phi_1) \cap O_i \subseteq U_i$ and 
$\vs \models (\keep(\phi_1) \setminus U_i )\bot (O_i \setminus U_i ) | U_i$. 
\ee  

In the final condition, $U_i$ can be any set. We note that a set $U_i$ satisfying the condition
can always be found. For, if we take $U_i = O_i$, then the condition states that 
 $\keep(\phi_1) \cap O_i \subseteq O_i$ and 
$M \models (\keep(\phi_1) \setminus O_i )\bot \emptyset | O_i$. 
Both parts of this statement are trivially true. In practice, we will want to 
choose $U_i$ to be as small as possible, since this will lead to stronger optimizations.%
\footnote{We remark also that since $(A\setminus C)\bot (B\setminus C)|C$ is equivalent to  
$A\bot B|C$, the independence condition could be more simply stated as 
$\keep(\phi_1) \bot O_i| U_i$. We work with the more complicated version because 
the algorithm for d-separation assumes disjoint sets.}

Note that $\phi$ is a subformula of itself, so in the domain of $\keep$. 
The following result says that satisfaction of $\phi$ is preserved when we marginalize 
to a superset of $\keep(\phi)$ for a relevance function $\keep$. 

\begin{theorem} 
Suppose that $\keep$ is a relevance function for $\phi$ with respect to epistemic variable structure $\vs$ 
and that $X$ is a set of variables with $\keep(\phi) \subseteq X \subseteq \dom{\vs}$. 
Then for all worlds $w$ of $\vs$, we have $\vs, w\models \phi$ iff $\vs\marg X, w \restrict X \models \phi$. 
\end{theorem} 

\begin{full} 
\begin{proof} 
We prove the result, for all epistemic variable structures $\vs$, by induction on the structure of $\phi$. 
Suppose $\keep(\phi) \subseteq X \subseteq \dom{\vs}$. 

For the case $\phi = p \in \Prop$, we have $\keep(\phi) = \{p\}$, so $p \in X$. 
Hence $\vs, w \models \phi$ iff $w(p) =1$ iff $(w\restrict X)(p) =1$  iff $\vs \marg X, w\restrict X \models \phi$, 
as required. 

In case $\phi = \phi_1 \land \phi_2$, we have $\keep(\phi) = \keep(\phi_1) \cup \keep(\phi_2)$, 
so $\keep(\phi_1) \subseteq X$ and $\keep(\phi_2) \subseteq X$. 
Hence 
\begin{align*} 
\vs, w \models \phi &  \text{ iff } \vs, w \models \phi_1 \text{ and } \vs, w \models \phi_2 \\ 
& \text{ iff } \vs\marg X, w \restrict X \models \phi_1 \text{ and } \vs\marg X, w \restrict X \models \phi_2 & \hfill \text{(by induction)}\\
& \text{ iff } \vs\marg X, w \restrict X \models \phi~.
\end{align*} 
The proof for the case $\phi = \neg \phi_1$ is similar. 

In case $\phi= K_i \phi_1$, we show that $\vs, w\models K_i \phi_1$ implies $\vs\marg X, w \restrict X \models K_i  \phi_1$, 
and the converse. Note that the equivalence relation $\sim'_i$ used in the semantics of the operator $K_i$ in  $\vs \marg X$ is
given by $v \sim'_i w$ if $v\restrict O_i \cap X = w \restrict O_i \cap X$.

For the implication from $\vs, w\models K_i \phi_1$ to $\vs\marg X, w \restrict X \models K_i\phi_1$, 
suppose that $\vs, w\models K_i \phi_1$. Let $w\restrict X  \sim'_i v$, where $v$ is a world of $\vs \marg X$. 
Then there exists a world $u$ of $\vs$ such that $u \restrict X = v$. 
We need to show that  $\vs\marg X, u \restrict X \models \phi_1$. 
Since $U_i \subseteq \keep(K_i \phi_1) \subseteq X$ and $U_i \subseteq O_i$, 
we have $U_i \subseteq X\cap O_i$. Hence, from $w\restrict X  \sim'_i u \restrict X$, 
we have $w \restrict U_i =  u \restrict U_i$. Thus, from 
$M \models (\keep(\phi_1) \setminus U_i )\bot (O_i \setminus U_i ) | U_i$, it follows 
that there exists a world $w'$ of $\vs$ such that 
$w'\restrict (O_i\setminus U_i) \cup U_i = w \restrict (O_i\setminus U_i) \cup U_i$ and 
$w'\restrict (\keep(\phi_1) \setminus U_i) \cup U_i = u\restrict (\keep(\phi_1) \setminus U_i) \cup U_i$. 
 Thus, $w'\restrict O_i=  w \restrict O_i$ and 
 $w' \restrict \keep(\phi_1)  = u \restrict \keep(\phi_1)$. Hence  
 \begin{align*} 
 \vs , w' \models \phi_1 ~~~~&  \text{iff }\vs\marg \keep(\phi_1)  , w'\restrict \keep(\phi_1) \models \phi_1&  \hfill \text{(by induction)}\\
  & \text{iff }\vs\marg \keep(\phi)   , u\restrict \keep(\phi) \models \phi_1 &  \hfill \text{(by $w' \restrict \keep(\phi_1)  = u \restrict \keep(\phi_1)$)} \\
    & \text{iff }\vs , u\models \phi_1 & \hfill \text{(by induction)}
\end{align*} 
Since $w'\restrict O_i=  w \restrict O_i$ and $\vs, w\models K_i \phi_1$, we have
$\vs, w'\models \phi_1$. Hence $\vs , u\models \phi_1$. 
By induction, $\vs\marg X, u \restrict X \models \phi_1$,  as required. 

Conversely, suppose that $\vs\marg X, w \restrict X \models K_i  \phi_1$. We show that 
$\vs, w\models K_i \phi_1$. For this, let $u$ be a world of $\vs$ with $w \restrict O_i = u \restrict O_i$. 
We need to show that $\vs, u\models \phi_1$. For this, note that it follows from 
$w \restrict O_i = u \restrict O_i$ that $w \restrict X \restrict O_i\cap X = u\restrict X \restrict O_i\cap X$, i.e.,  
$w \restrict X  \sim'_i  u\restrict X$. 
Since $\vs\marg X, w \restrict X \models K_i  \phi_1$, we have that 
$\vs\marg X, u \restrict X \models  \phi_1$. Since $\keep(\phi_1) \subseteq X$, we have, by induction, that 
$\vs, u  \models  \phi_1$, as required. 
\end{proof} 
\end{full} 

{\bf Computing $\keep(\phi)$:}  The definition of $\keep$ provides a recursive
definition by which $\keep (\phi)$ can be calculated, with the exception that the 
case $\keep(K_i(\phi)) = U_i \cup \keep(\phi)$ allows for a choice of the set $U_i$, subject to 
the conditions $\keep(\phi) \cap O_i \subseteq U_i$ and 
$\vs \models (\keep(\phi) \setminus U_i )\bot (O_i \setminus U_i ) | U_i$. 
When the worlds of $\vs$ are represented by a structured relational model $M$,  
we show how to construct the \emph{minimal} set $U_i$ satisfying the stronger conditions
that  $\keep(\phi) \cap O_i \subseteq U_i$ and  $U_i$ d-separates  $\keep(\phi) \setminus U_i $ from  $O_i \setminus U_i $
in the directed graph $G$ associated with $M$.

Note $(\keep(\phi) \setminus U_i) \cup  (O_i \setminus U_i) \cup U_i= \keep(\phi) \cup O_i$
for any set $U_i$. Thus, the d-separation properties we
are interested in are computed in the moralized graph $H = (G_{\ancestors{O_i\cup \keep(\phi)}})^m$, 
which is independent of $U_i$. 
Let $U$ be the set of vertices $v\in O_i$ such that there exists 
a path in $H$
from a vertex $u \in \keep(\phi)\setminus O_i$ to $v$, 
with $v$ the first vertex on that path that is in $O_i$. 
The set $U$ can be constructed in linear time by a depth first search from $\keep(\phi)\setminus O_i$. 
Take $W = U \cup (\keep(\phi) \cap O_i)$.

\begin{proposition} 
$W$ is the smallest set satisfying the strengthened conditions for $U_i$. 
\end{proposition} 

\begin{full} 
\begin{proof} 
We first show that $W$ satisfies the conditions for $U_i$. Clearly $\keep(\phi) \cap O_i \subseteq W$. 
We show that $\keep(\phi) \setminus W$ is d-separated from $O_i \setminus W $ by $W$.  
Let $u= u_0, u_1, \ldots, u_n =v$ be a path in   $H$
from $u \in \keep(\phi) \setminus W$ to $O_i \setminus W $. 
Since $\keep(\phi)\cap O_i \subseteq W$, we have $u \not \in O_i$. 
But then the path must cross an edge from the exterior of $O_i$ into $O_i$, 
and the endpoint of that edge is in $W$, by definition. This shows that there is no 
path from $\keep(\phi) \setminus W$ to $O_i \setminus W$ that avoids $W$. 
Thus, $W$ satisfies the conditions for $U_i$. 

To show that $W$ is the minimal such set, let $U_i$ be any set satisfying the conditions, 
and suppose that $W\not \subseteq U_i$. Then there is a vertex $v\in  W\setminus U_i$. 
Since $W\subseteq O_i$, we have $v\in O_i \setminus U_i$. 
We cannot have $ v \in \keep(\phi) \cap O_i$, since then also $v\in U_i$. 
Thus, by definition of $W$,  there exists a path in $H$ 
from $v$ to some node in $\keep(\phi) \setminus O_i$, with all vertices after $v$ not in $O_i$, 
hence also not in $U_i$ since $U_i \subseteq O_i$.
Note $\keep(\phi) \setminus O_i = \keep(\phi) \setminus U_i$ since $\keep(\phi)\cap O_i \subseteq U_i$. 
Thus, $\keep(\phi)\cap O_i \subseteq U_i$ is not d-separated from $O_i \setminus U_i$ by
$U_i$, contradicting the assumptions on $U_i$. This shows that $W\subseteq U_i$. 
\end{proof} 
\end{full} 

\subsection{Equalities} \label{sec:equalities} 
Unfolding a program into a structured model tends to create a large number of 
timed variable instances whose associated value represents an equality between
two variables. Such instances can be eliminated by a simple transformation of the 
structured model.

For an assignment $\alpha$ with domain $V$, define $\alpha[y/x]$ to be the 
assignment $\alpha'$ with domain $(V\setminus\{x\})\cup \{y\}$ with 
$\alpha(y) = \alpha'(x)$ and $\alpha\restrict (\dom{s} \setminus \{x\} )= \alpha'\restrict (\dom{s} \setminus \{x\})$.

For a relational value $s$ and variables $x,y$ with $x \in \dom{s}$ and 
$y \not \in \dom{s}$, define $s[y/x]$ to be the relational value $t$ 
with $\dom{t} = (\dom{s} \setminus \{x\})\cup \{y\}$, 
consisting of all assignments $\alpha[y/x]$ for $\alpha \in s$. 
Intuitively, this is simply the relation $s$ with variable $x$ renamed to $y$. 

We extend this definition to structured relational models $M = (V,E,\relss)$ with $x,y\in V$, 
by defining $M[y/x] = (V', E', \relss')$ with 
$V' = V\setminus \{x\}$, and $E'=  E\cap (V' \times V')$, and 
$\relss = \{s'_v~|~v\in V'\}$, where $s'_v = s[y/x]$. 
In the following result, we write $\delta_{x,y}$ for the set of 
assignments $\alpha$ with domain $\{x,y\}$ and $\alpha(x) = \alpha(y)$. 

 \begin{proposition} 
Suppose that $M = (V,E,\relss)$ is a structured relational model with $x,y\in V$, and $\val{x} = \val{y}$, and 
$\parents{y} = \{x\} $ and $s_y = \delta_{x,y}$. Let $M[y/x] = (V', E', \relss')$. 
Then  $\comb \relss' = (\comb \relss) \marg V'$. 
 \end{proposition}

The definition furthermore extends to epistemic models 
$\vs = (A,O,V)$ with worlds represented by a 
structured relational model $M = (V, E, \relss)$. Let $M[y/x] = (V', E', \relss')$.  We define $\vs[x/y] = (A',O',V')$ where
$O' = \{O'_i\}_{i \in \agts}$ where $O'_i = O_i \cup \{y~|~x\in O_i\}$ for each $i \in \agts$, 
and $A' = \comb \relss'$. Note that $O'_i$ additionally makes variable $y$ visible to agent $i$ if 
$x$ was visible to $i$, in case this variable was not originally visible. 
 
 \begin{proposition} 
If $\val{x} = \val{y}$, and  $\parents{y} = \{x\} $ and $s_y = \{(x:a,y:a)~|~ a \in \val{x}\}$
then $\vs, \alpha \models \phi$ iff $\vs[y/x], \alpha[y/x] \models \phi[y/x]$. 
 \end{proposition} 
\begin{full} 
\end{full} 

\subsection{Algorithm} 

The overall  optimized procedure for model checking  that we obtain from the above results 
uses the following steps: 
\be
\item We first unfold a program representation of the model into a structured relational model with symbolically represented values
and transform the query into a form that uses the timed instances variables in place of the 
original variables. This can be done in a way that builds in the equality optimization of 
Section~\ref{sec:equalities}. We expand on this step in Section~\ref{sec:symbeval}. 
\item We compute $\keep(\phi)$ using the algorithm in Section~\ref{sec:elimovars}. 
\item We compute a symbolic representation of $\vs\restrict \keep(\phi)$, 
using the leaf node elimination optimization. 
\item We compute $\vs\restrict \keep(\phi) \models \phi$ in this representation
using a symbolic model checking algorithm.
\ee

\begin{full} 
\section{Example} \label{sec:dcopt} 

In the present section, we illustrate this procedure on the 
Dining cryptographers protocol. 

Figure~\ref{fig:optgraph} indicates the dependency graph that remains after we have 
applied the optimization procedure to the Dining cryptographers problem.  
We consider the formula 
$$\phi = ( \neg \paid_0 \rimp K_0(\neg \paid_1 \land \neg \paid_2) \lor (K_0 (\paid_1 \lor \paid_1) \land  \neg K_0 \paid_1 \land \neg K_0 \paid_2)$$
evaluated at time 3. 
Transforming to timed form, this is 
$$\phi_1 = ( \neg \paid^3_0 \rimp K_{0}(\neg \paid^3_1 \land \neg \paid^3_2) \lor (K_0 (\paid^3_1 \lor \paid^3_1) \land  \neg K_0 \paid^3_1 \land \neg K_0 \paid^3_2)$$
The set of observable variables  $O_0$ used for the operator $K_0$ in this formula 
is the set of all variables inside the rectangles in Figure~\ref{fig:initgraph}. 

The result of applying the equality optimization to the  model is 
depicted in Figure~\ref{fig:graph-eq}. 
The resulting formula is 
$$\phi_2 = ( \neg \paid^0_0 \rimp K_{0}(\neg \paid^0_1 \land \neg \paid^0_2) \lor (K_0 (\paid^0_1 \lor \paid^0_1) \land  \neg K_0 \paid^0_1 \land \neg K_0 \paid^0_2)$$

\begin{figure} 
\centerline{\includegraphics[height=3in]{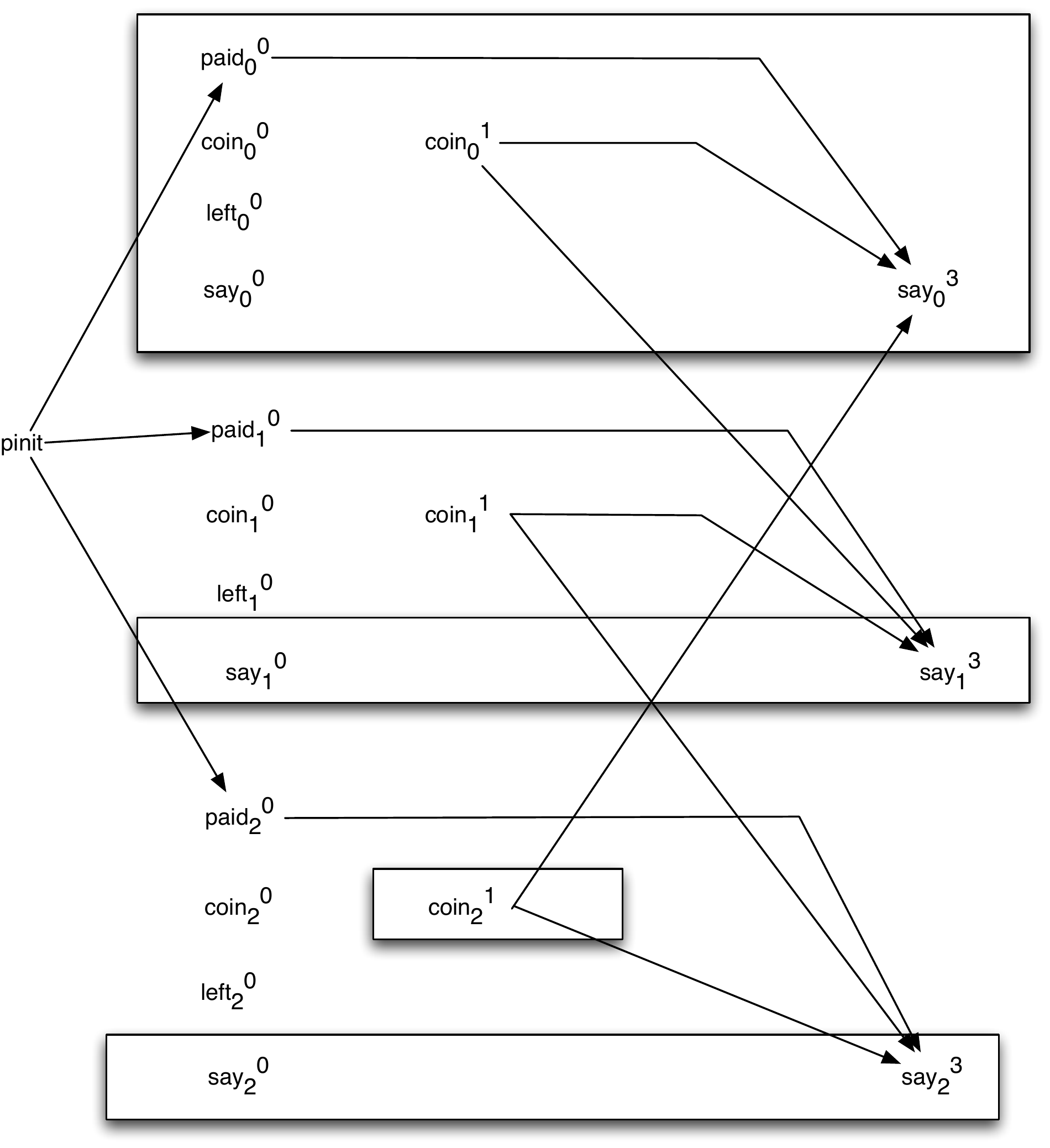}} 
\caption{Depency graph after equality optimization \label{fig:graph-eq}} 
\end{figure}

To construct the moralized graph, we add edges between all vertices in 
the sets $\{\paid^2_i, \coin^2_i, \leftcoin^2_i\}$ for $i=0,1,2$, 
and replace all directed edges with undirected edges. 
The result is depicted in Figure~\ref{fig:graph-moral}.

\begin{figure} 
\centerline{\includegraphics[height=3in]{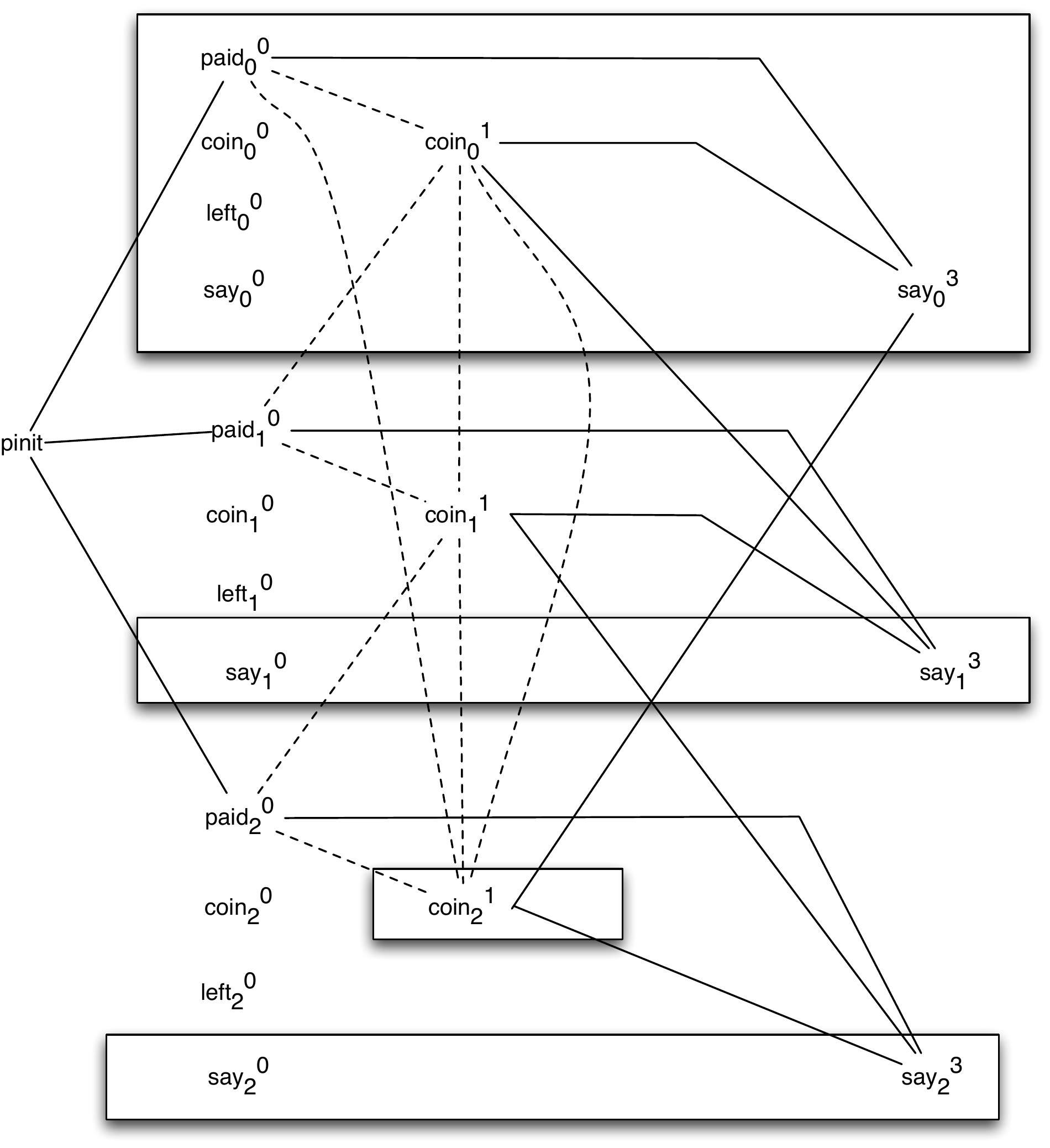}} 
\caption{Moralised graph\label{fig:graph-moral}} 
\end{figure} 

For the computation of $\keep{\phi}$,  we note that the 
variables in the scope of the knowledge operator are 
$\paid_1^0$ and $\paid_2^0$. The vertices at the 
outer boundary to the observable variables from these
vertices are $\pinit, \paid_1^0, \paid_2^0, \coin^1_1$. 
The observable variables reachable in one step from this outer  boundary
are $\paid_0^0,  \coin^1_0, \coin^1_1, \say^3_1, \say^3_2$. 
Thus, we compute 
$$\keep(\phi_2) = \{ \paid_0^0, \paid_1^0,\paid_2^0 \coin^1_0, \coin^1_1, \say^3_1, \say^3_2\}$$ 
All other variables can be eliminated using the variable  elimination algorithm. 
As a first step in this process, we can delete leaf nodes not in 
$\keep(\phi_2)$ (and recursively, any fresh leaf nodes not in $\keep(\phi_2)$  resulting from such deletions.) 
This step enables deletion of variables  $\say^3_0$ and $\coin_i^0, \leftcoin^0_i, say^0_i$ for $i = 0,1,2$. 
The graph resulting from these deletions in 
depicted in Figure~\ref{fig:optgraph}.

\begin{figure} 
\centerline{
\includegraphics[height=3in]{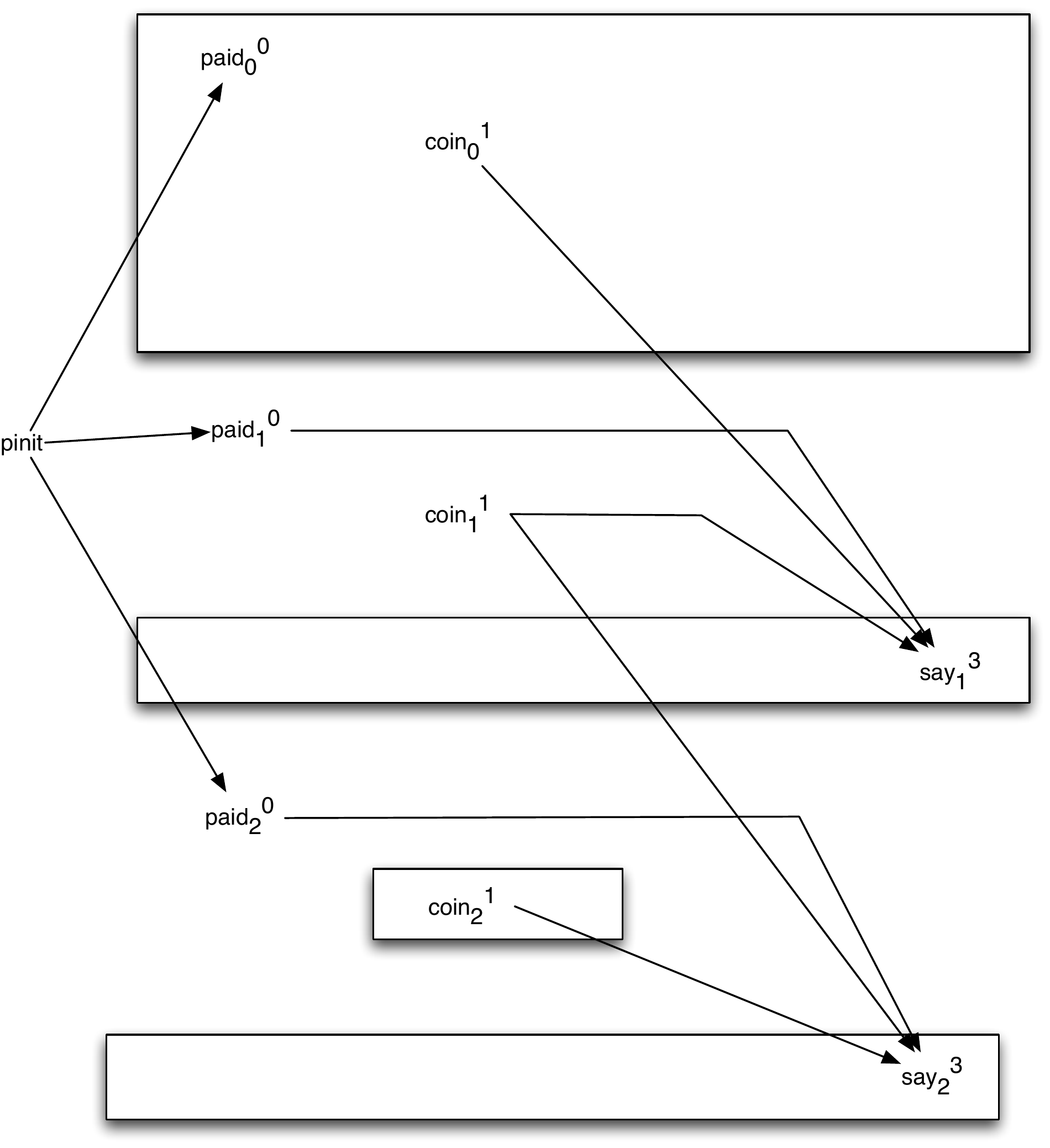}
}
\caption{Dining Cryptographers Dependency graph after optimization\label{fig:optgraph}} 
\end{figure} 

From the point of model checking complexity, we expect that the simplification of the 
dependency graph will result in significant improved performance of the model checking computation. 
For $n$ cryptographers,  the initial dependency graph (Figure~\ref{fig:initgraph} for $n=3$) has $16 n$ variables, 
i.e., $48$ variables in case $n=3$. The algorithm of van der Meyden and Su \cite{MS} would construct a BDD with over
$12+4n$  variables in general, and, as show in Figure~\ref{fig:meydensu},  with 24 variables in case $n=3$. 
However, the algorithm uses an intermediate BDD representation of the transition relation of the 
protocol that requires $8n$ variables. Instead, the optimization 
approach developed here computes a BDD over just 9 variables in case $n=3$ and 
$3n$ variables in general. The actual model checking computation combines BDD's associated with each node
to construct a BDD over the same number of variables. 
Since in practice, BDD algorithms work for numbers of variables in the order of 100-200, 
these reductions of the constant factor can have a significant impact on the scale of the 
problems that can be solved. 

\begin{figure} 
\centerline{
\includegraphics[height=10cm]{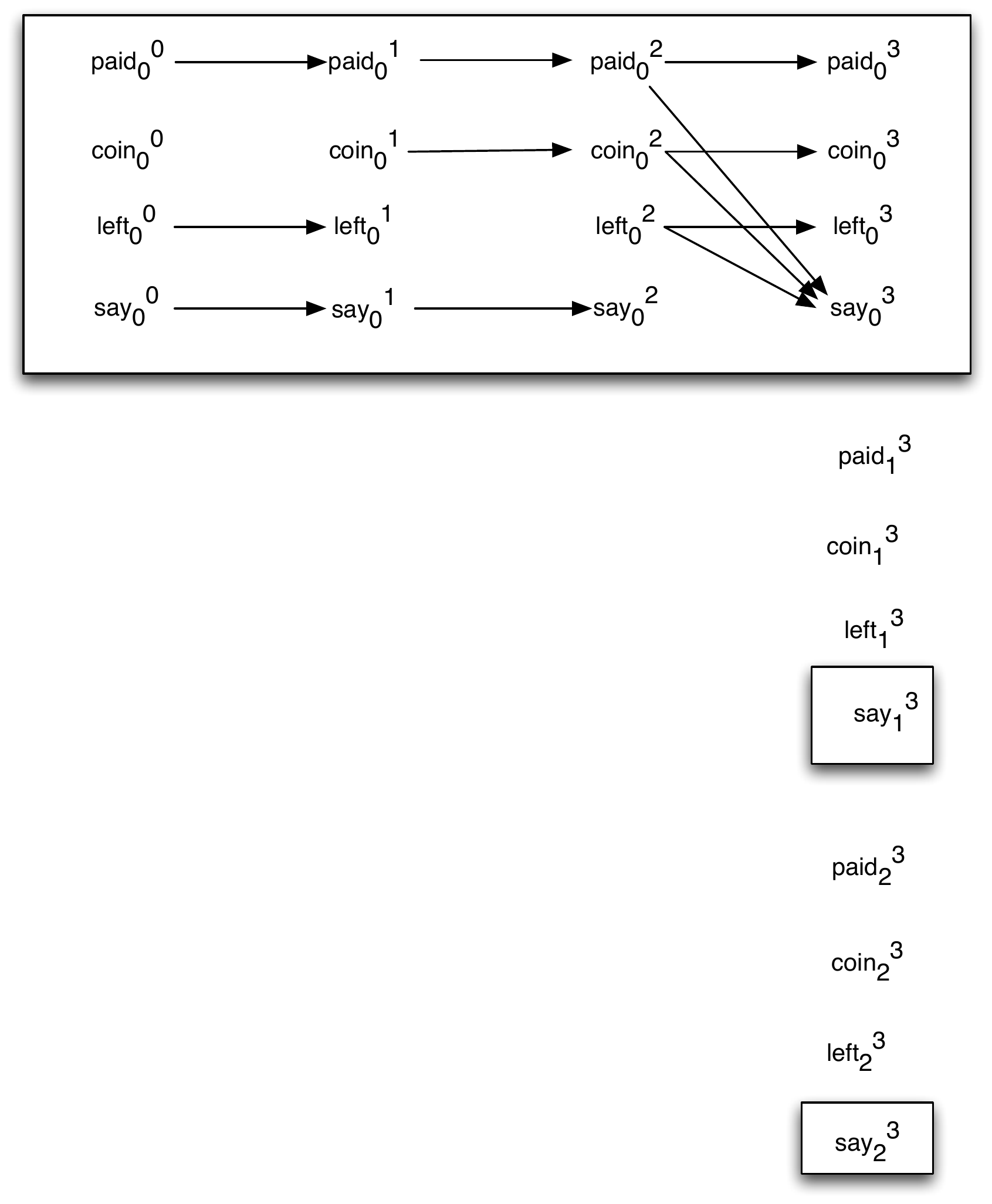}
}
\caption{Timed-variables used in algorithm of van der Meyden and Su. \label{fig:meydensu}} 
\end{figure} 

\end{full} 

\section{From Programs to Dags} \label{sec:symbeval}

We have developed an implementation of the above ideas
as an extension of the epistemic model checker MCK \cite{mck}.
We sketch the  implementation in this section.

We apply the conditional independence optimization  described above on a 
structured model derived from the system, in which values are represented symbolically as formulas. 
Given a formula to be model checked, we derive a structured model over  a smaller set of vertices using the conditional independence 
optimization.  Rather than producing the timed variable dag 
as in the discussion of the Dining Cryptographers example above, and then applying the equational 
optimization, the initial structured model  is obtained by means of  a symbolic execution that builds in 
the equational optimization. This symbolic execution proceeds as follows. 

\newcommand{\vind}{\gamma}  
\newcommand{\nat}{\bf N} %
\newcommand{\modupdate}{\mu}

For a set of variables $\Vars$, define an 
\emph{indexing} of $\Vars$ to be a mapping $\vind: \Vars \rightarrow \nat$. 
If $\vind$ is an indexing, and $e$ is an expression, 
we define the the expression $e/\vind$, which interprets $e$ with 
respect to indexed instances of the variables, by 
replacing each occurrence of a variable $v$ in $e$ by the indexed variable 
$v^{\vind(v)}$. Intuitively, $v^{k}$ represents the $k$-th value taken on by 
variable $v$ during the running of the program. (Note that this differs
from the timed variable $v_k$, which represents the value of the variable at time $k$.)

Consider a  system $\sys = (J,I,Q)$ 
with joint protocol $J= P_1 ~|| ~\ldots ~||~ P_n  ~\Delta~ C_E$. 
We can, for each time $t$ up to the maximal running time $N$ of $J$, 
obtain the code at time $t$, denoted $C_t = C^t_1; \ldots C^t_n; C_E$, 
where the $t$-th atomic statement in each $P_i$ is $\atl C^t_i\atr$. 
(At this step, we use the fact that the agent protocols are straightline.) 
We construct a sequence of structured models 
$M_0, \ldots , M_N$, and a sequence of indexings $\vind_0, \ldots \vind_N$ of $\Vars$, 
as follows.

To represent the initialization condition, we use a variable $v_{init}$  with frame
$\Omega_{v_{init}}$ equal to the set of assignments over $vars(I)$. (Under our simplifying assumptions, 
this is a set of assignments to boolean variables; all indexed variables other than $v_{init}$ are boolean.)

The initial structured model $M_0  = \langle V_0, E_0,\relss\rangle$ has 
$V_0 = \{ v^0~|~ v\in \Vars\}\cup \{v_{init}\}$ and $E_0 = \{(v_{init},v^0)~|~v\in vars(I)\}$. 
Write $s_u= (A_u,V_u)$ for the values $\relss = \{s_u\}_{u \in V_0} $.  
The domains of these values are given by 
$V_{v_{init}} = \{v_{init}\}$
and 
$V_{v^0} =  \{v^0 \}$ if $v\not \in vars(I)$ and 
$V_{v^0} =  \{v_{init}, v^0 \}$ otherwise. 
The relations in these values are symbolically represented 
$A_{v_{init}} = I$, and 
$A_{v^0} = True$ if $v\not \in vars(I)$, otherwise 
by $v^0 = v$.  
The indexing $\vind_0$ is the \emph{initial indexing}, which has $\vind_0(v) = 0 $ for all $v\in \Vars$. 

Given model $M_i$, indexing $\vind_i$ and code $C_i$, we obtain the 
next  model and indexing in the sequence as $(M_{i+1},\vind_{i+1}) = \modupdate(M_i,\vind_i,C_i)$, 
where the function $\modupdate$ is defined by 
$$ \modupdate(M, \vind, \epsilon) = (M, \vind)$$ 
and 
$$ \modupdate(\langle V,E, \relss\rangle, \vind, (b;C)) = (\langle V',E', \relss'\rangle, \vind',C)$$ 
where, if $b$ is the assignment $v:=e$ we have 
$$
\begin{array}{l}
V' = V\cup \{v_{\vind(v)+1}\} \\ 
E' = E\cup \{ (u_{\vind(u)},v_{\vind(v)+1}) ~|~u \in vars(e)\} \\
\relss' = \relss \cup \{s_{v_{\vind(v)+1}}\} \\ 
\vind' = \vind[\vind(v)+1/v]
\end{array} 
$$ 
with the relation of $s_{v_{\vind(v)+1}}$ symbolically represented by the formula $v_{\vind(v)+1} \dimp e/\vind$. 
That is, the function $\modupdate$ processes an 
assignment statement $v:=e$ by interpreting $e$ with respect to $\vind$, 
creating a new vertex $v_{\vind(v)+1}$ with parents the variables in this interpretation 
$e/\vind$ and a value that describes how $v_{\vind(v)+1}$ is calculated from 
its parents. 

In case $b$ is the randomization statement $rand(v)$, we take 
$$\begin{array}{l}
V' = V\cup \{v_{\vind(v)+1}\} \\ 
E' = E \\
\relss' = \relss \cup \{s_{v_{\vind(v)+1}}\} \\ 
\vind' = \vind[\vind(v)+1/v]
\end{array} 
$$ 
with the relation of $s_{v_{\vind(v)+1}}$ symbolically represented by the formula $True$. 

The sequence of indexings $\vind_0\ldots v_N$ relates timed
variables to indexed variables, by the mapping $v_t \mapsto \gamma_t(v)$.  
Via this mapping, the structured model $M_N$ represents the 
worlds of the epistemic variable structure $\vs_N(\sys)$. 

In particular,  to evaluate an  epistemic formula $\phi$ at time $N$, we 
work with the model $M_N$ and  interpret each variable
$v$ of $\phi$ as the indexed variable $\vind_N(v)$. 
In order to determine $\keep(\phi)$, where $\phi$ is a formula to be evaluated at time $N$, we use the 
sets $O_i = \{v^t~|~v\in Q_i,~0\leq t\leq \gamma_N(v)\}$ of images of observable timed
variables.    

After constructing the structured model $M_N$ and 
computing the set of relevant variables $\keep(\phi)$,  
we compute $M_N \restrict \keep(\phi)$ using the 
leaf node optimization. This is again a structured 
model. Since the values in $M_N$ are formulas, 
$M_N \restrict \keep(\phi)$ is represented as
a formula of quantified boolean logic. 
We then process this formula from the leaves to the
root to obtain a binary decision diagram 
representing this QBF formula -- the variables in 
this representation are the variables in $\keep(\phi)$. 
The assignments represented by this 
binary decision diagram are the worlds of a
Kripke structure. 

The observable variables yield binary relations over these
worlds, defined by $ w\sim_i w'$
if for all variables $v\in O_i$, we have 
$w(v^{\vind_N(v)}) = w'(v^{\vind_N(v)})$. 
These relations correspond to an agent with perfect 
perfect recall observing variables $O_i$. 
The relations can similarly be represented by
binary decision diagrams. This yields a symbolic representation 
of the Kripke structure using binary decision diagrams. 

A standard symbolic evaluation procedure for modal logic can 
then be used to compute a BDD representing the set of worlds 
where $\neg\phi$ holds. We check this for emptiness to 
decide $M \models \phi$. 

\section{Experimental Results} \label{sec:results} 

In the present section, we describe the results of a number of experiments designed to 
evaluate the performance of epistemic model checking using the conditional 
independence optimization, in comparison with the existing implementation in 
MCK. (Since MCK remains the only  symbolic epistemic model checker that deals with 
perfect recall knowledge, there are no other systems to compare to.) 
All experiments were conducted on  an Intel 2.8 GHz Intel Core i5 
processor with 8 GB 1600 MHz DDR3 memory running Mac OSX 10.10.

The experiments conducted are based on a number of examples of epistemic 
model checking applications that have previously been considered in the 
literature. Most concern security protocols. Each experiment 
scales according to a single numerical parameter, and we measure
the running time of model checking a formula as a function of this parameter. 
\begin{full}
Details of the protocols and the way that they are represented in the MCK 
scripting language when using the conditional independence optimization 
are presented in Appendix~\ref{appendix}.%
\footnote{The scripts are presented using a version of MCK's language 
under development for an upcoming release  -- this is more elegant than earlier versions, 
and we developed the implementation of the optimization  to work with the new language. 
The performance of model checking is sensitive to the encoding of a script to the
quantified boolean formulas used by the model checking algorithms. 
Because the unoptimized model checker had not yet been fully adapted to the new language
at the time we conducted this work, 
we used alternate but logically equivalent scripts for the running times of the non-optimized computation. 
These scripts were chosen so as to minimize the running times for the non-optimized 
version, so as to best advantage the non-optimized version in the competition. 
(Even with this advantage to the non-optimized version, the optimization generally wins.)  
} 
\end{full}

The performance of the conditional independence optimization 
depends on the extent to which it is able to reduce the number of
variables that need to be handled in the ultimate BDD calculation
that implements model checking. Theoretically, in the worst case, 
there is no reduction in the number of variables. 
We have therefore deliberately chosen some examples in which the
reduction is realised in order to demonstrate its power when it 
applies.   However, the examples are realistic in that they derive from 
prior, independently motivated work.

Except where indicated, the unoptimized model checking algorithm against which 
we compare is that invoked by the construct {\tt spec\_spr\_xn} in the MCK scripting language, 
which operates as already described above. (We refer to this algorithm as {\tt xn} in 
legends, and the algorithm using conditional independence optimization is referenced as {\tt ci}.) 
The results demonstrate both significant speedups of as large as four orders of magnitude, 
as well as a significant increase in the scale of problem that can be handled in 
a give amount of time. 

Due to nondeterminism in the underlying CUDD package \cite{cudd} used by 
MCK for binary decision diagram computations, the running times 
can show significant variance from run to run, with some runs 
taking very large amounts of time.  We have dealt with this variance 
by concentrating on the \emph{minimal} running time obtained over three runs of the 
experiments. This form of aggregation can be justified as equivalent to 
the running time obtained when running three copies of the computation in 
parallel and taking the answer from the first to complete. 

Even with this allowance, the plot of running times 
as we scale the experiments can be very jagged on larger instances.
The problem tends to affect the unoptimized running times more than 
the running times using the conditional independence optimization, which 
generally give smooth curves.  We believe this is due to memory placement 
effects on larger BDD's, and because the BDD's for 
the unoptimized running times reach the critical size significantly earlier. We expect that a similar 
phenomenom will occur with the optimized version once this critical BDD size is reached.

 \subsection{Dining Cryptographers} 
 
Our first example is the Dining Cryptographers protocol \cite{chaum}, as discussed in Section~\ref{sec:exampledc}. 
This scales by the number $n$ of agents; the number of state variables is $O(n)$, 
and the protocol runs for 3 steps. The initial condition needs to say that at most one
of the agents paid -- this is done by means of a formula of size $O(n^2)$. The rest of the script scales linearly. 
The formula in all instances  states that at time 3, agent $C0$ either knows that 
no agent pays, knows that C0 is the payer, or knows that 
one of the other cryptographers is the payer, but does not know which. 
This involves $O(n)$ atomic propositions, and is of linear size in $n$.

Performance results for model checking the Dining Cryptographers protocol running on a ring with $n$ agents are
shown in Table~\ref{fig:dc:results-xn}. There is a rapid blowup as the number of agents is increased: 12 agents 
already takes over 46 minutes (2775 seconds). 

\begin{table} \centerline{
\begin{tabular}[b]{|c || c | c |   c | c |   c | c |   c | c |   c | c |   c | c |   }  
\hline
$n$ &  3 & 4& 5& 6& 7& 8& 9& 10& 11& 12 \\ 
\hline 
xn & 0.1&  0.16&  0.37&  0.65&  1.34&  2.52&  37.92&  33.5&  846.19&  2775.4\\
\hline 
\end{tabular}}
\caption{\label{fig:dc:results-xn} Dining Cryptographers experiments, unoptimized running times (s) } 
\end{table} 

\begin{figure} 

\centerline{\includegraphics[height=9cm]{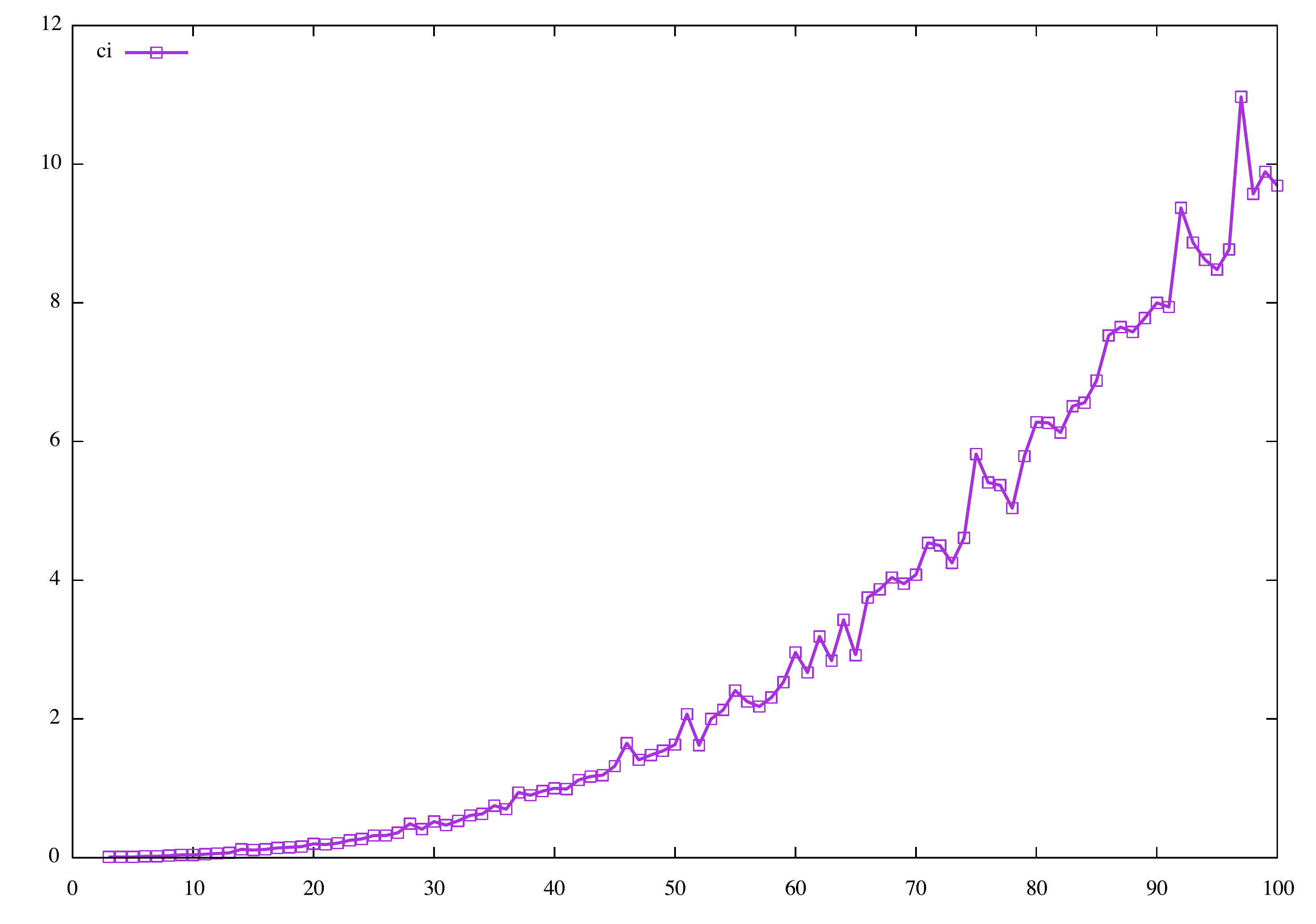}  }

\caption{\label{fig:dc:results} Dining Cryptographers experiments, optimized  running times (s) } 
\end{figure}

By contrast, applying the conditional independence optimization, 
model checking is significantly more efficient, as shown by the plot in 
Figure~\ref{fig:dc:results}. The case of 12 agents is handled in 0.05 seconds, 
and 100 agents are handled in 9.69 seconds.  

\subsection{One-time Pad} 

The next example concerns message transmission using one-time pad encryption in the presence of an eavesdropper.
Each instance has three agents (Alice, who sends an encrypted message to Bob,
and Eve, who taps the wire). We scale the example by the length of the message, which is sent one bit at a time.  
For a message of length $n$, states have $O(n)$ variables. 
The protocol runs $2n$ steps, two for each bit.  
The formula is evaluated at time $2n$, and 
says that Eve does not learn the value of the first bit. 

For this example, we found that the best performance for the unoptimized version was
obtained using MCK version 0.5.1, which used a different encoding from more recent versions. 
Performance of model checking is shown in Table~\ref{fig:results-otp}. 

The running times for the optimized version grow very slowly (the numbers show
a step-like behaviour due to rounding). 
Intuitively, the conditional independence optimization 
detects in this example that the first bit and the others are independent, and uses this 
to optimize the model checking computation. This means that for all $n$, the ultimate BDD model checking computation is 
performed on the same model for all $n$, and the primary running time cost lies in the generation of the 
dependence graph, and its analysis, that precedes the BDD computation. 
On the other hand, the unoptimized (xn) model checking running times show significant growth, with a large spike
towards the end, where the speedup obtained from the optimization is over 10,000 times. 

\begin{table} \centerline{
\begin{tabular}[b]{|c || c | c |   c | c |  }  
\hline 
$n$ & ci (s) & xn (s) & xn/ci \\
\hline  
3& 0.01& 0.03& 3\\
4& 0.01& 0.08& 8\\
5& 0.01& 0.10& 10\\
6& 0.01& 0.18& 18\\
7& 0.01& 0.27& 27\\
8& 0.01& 0.35& 35\\
9& 0.01& 0.52& 52\\
10& 0.02& 0.52& 26\\
11& 0.02& 1.25& 63\\
12& 0.02& 1.38& 69\\
13& 0.02& 2.04& 102\\
14& 0.02& 2.19& 110\\
15& 0.02& 3.98& 199\\
16& 0.03& 5.50& 183\\
17& 0.03& 5.01& 167\\
18& 0.03& 5.47& 182\\
19& 0.03& 7.24& 241\\
20& 0.04& 9.71& 243\\
21& 0.04& 8.42& 211\\
22& 0.04& 8.82& 221\\
23& 0.04& 11.10& 278\\
24& 0.04& 17.88& 447\\
25& 0.04& 36.68& 917\\
26& 0.04& 33.26& 832\\
27& 0.05& 23.60& 472\\
28& 0.05& 34.88& 698\\
29& 0.05& 99.50& 1990\\
30& 0.05& 50.10& 1002\\
31& 0.05& 75.13& 1503\\
32& 0.05& 67.37& 1347\\
33& 0.06& 97.23& 1621\\
34& 0.06& 184.19& 3070\\
35& 0.06& 89.47& 1491\\
36& 0.07& 131.74& 1882\\
37& 0.07& 164.76& 2354\\
38& 0.07& 259.48& 3707\\
39& 0.07& 275.87& 3941\\
40& 0.07& 749.88& 10713\\
\hline 
\end{tabular}}
\caption{\label{fig:results-otp} One-time pad protocol, optimized  and unoptimized running times, and speedup ratio, ``single-bit'' formula} 
\end{table}

\subsection{Oblivious Transfer} 

The next example concerns an oblivious transfer protocol due to Rivest \cite{rivest99}, 
which allows Bob to learn exactly one of Alice's two messages $m_0,m_1$, of his choice, without Alice knowing which
message was chosen by Bob.  
Each instance has two agents, and we scale by the length of the message. 
For a message of length $n$, states have $O(n)$ variables. 
We consider two formulas for this protocol. 
Both are evaluated at time 3 in all instances. 

The first formula says that if Bob chose to receive message $m_1$, then he does not learn the \emph{first} bit of $m_0$. 
The running times for model checking this formula are 
given in Table~\ref{fig:results-bbcs-single}. In this example, the 
conditional independence optimization gives a significant speedup, 
in the range of one to two orders of magnitude (more precisely, 12 to 221) 
improvement on the inputs considered, and increasing as the scale of the problem increases. 

\begin{table} \centerline{
\begin{tabular}[b]{|c || c | c |   c | c |  }  
\hline 
$n$ & ci (s) & xn (s) & xn/ci \\
\hline  
3& 0.02& 0.24& 12\\ 
4& 0.03& 0.52& 17\\
5& 0.05& 0.90& 18\\
6& 0.07& 1.80& 26\\
7& 0.11& 2.24& 20\\
8& 0.14& 3.54& 25\\
9& 0.15& 4.97& 33\\
10& 0.16& 7.20& 45\\
11& 0.21& 13.08& 62\\
12& 0.26& 16.68& 64\\
13& 0.32& 32.72& 102\\
14& 0.39& 62.08& 159\\
15& 0.43& 50.95& 118\\
16& 0.50& 36.73& 73\\
17& 0.60& 38.36& 64\\
18& 0.71& 69.27& 98\\
19& 0.77& 170.16& 221\\
20& 1.09& 148.56& 136\\
\hline 
\end{tabular}}
\caption{\label{fig:results-bbcs-single} Rivest Oblivious Transfer Protocol, optimized  and unoptimized running times, and speedup ratio, ``single-bit'' formula} 
\end{table} 

\begin{figure} 
\centerline{\includegraphics[height=14cm]{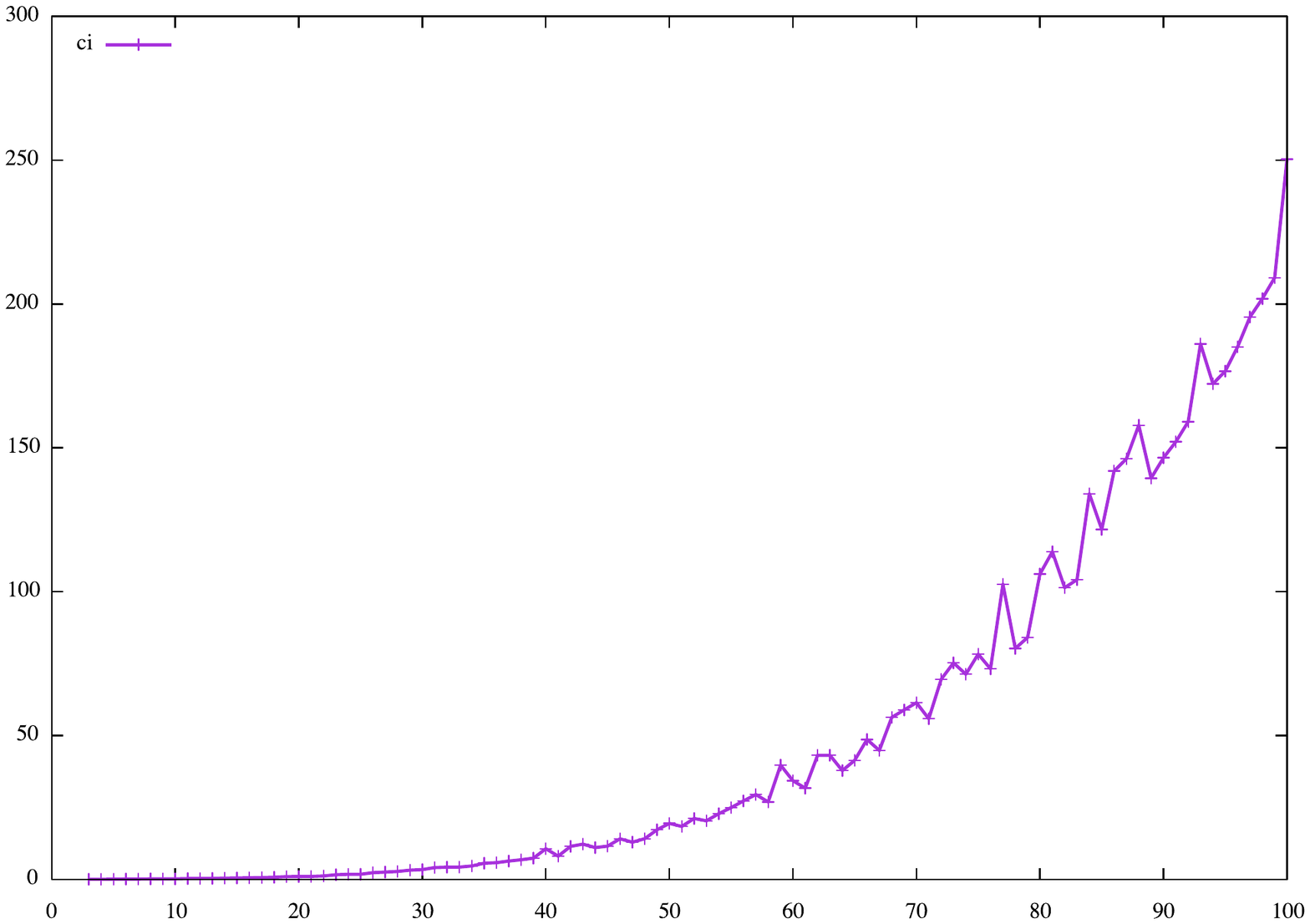}  }

\caption{Oblivious transfer protocol experiments, optimized running times, ``single-bit'' formula
\label{fig:bbcs-ci-only}} 
\end{figure}

Running just the optimized version on larger instances, we
obtain the plot shown in Figure~\ref{fig:bbcs-ci-only}. 
This shows that the optimization allows us to handle significantly 
larger instances: up to 97 agents can be handled in under 200 seconds, 
compared with 19 agents in 170 seconds unoptimized. 

An example in which the optimization does
not always yield a performance improvement 
arises when we change the formula model checked in this 
example to one that states that if Bob chose to receive $m_1$, then he does
not learn the value of \emph{any} bit of $m_0$. The running times are shown in Table~\ref{fig:results-bbcs-all}.

\begin{table} \centerline{
\begin{tabular}[b]{|c || c | c |   c | c |  }  
\hline 
$n$ & ci (s) & xn (s) & xn/ci \\
\hline 
3& 0.03& 0.25& 8.3\\
4& 0.05& 0.51& 10.2\\
5& 0.12& 0.86& 7.2\\ 
6& 0.15& 1.58& 10.5\\
7& 0.25& 2.84& 11.4\\
8& 0.42& 3.52& 8.4\\
9& 0.50& 5.11& 10.2\\
10& 0.55& 7.79& 14.2\\
11& 1.18& 13.07& 11.1\\
12& 3.72& 14.63& 3.9\\
13& 5.20& 39.74& 7.6\\
14& 7.13& 48.64& 6.8\\
15& 4.91& 56.62& 11.5\\
16& 20.16& 38.09& 1.9\\
17& 32.95& 42.40& 1.3\\
18& 174.96& 86.81& 0.5\\
19& 229.85& 96.86& 0.4\\
20& 342.40& 184.08& 0.5\\
\hline 
\end{tabular}}
\caption{\label{fig:results-bbcs-all} Rivest Oblivious Transfer Protocol, optimized and unoptimized  running times, and speedup ratio, ``all bits'' formula} 
\end{table} 

Here, the optimization initially gives a speedup of roughly one order of magnitude, but 
on the three largest examples, the performance of the unoptimized algorithm is better
by a factor of two. The lower size of the initial speedup, compared to the first formula,  can be explained from the
fact that are obviously fewer variables that are independent of the second formula, since the 
formula itself contains more variables. 
(The ``all bits'' formula contains $O(n)$ rather than just one variable explicitly, but recall that 
knowledge operators implicitly  introduce more variables, 
so the ``first bit'' formula implicitly has $O(n)$ variables.) 
It is not immediately clear exactly what accounts for the switchover. 

\subsection{Message Transmission} 

The next example concerns the transmission of a single bit message 
across a channel that is guaranteed to deliver it, but with uncertain delay
This example has two agents Alice and Bob , and runs for $n+1$ steps, where $n$ is the 
maximum delay.  States have $O(n)$ variables. The formula considered  
states at time $n+1$ that Alice knows that Bob knows ... (nested five levels) that the message  has arrived. 
Because of the nesting, the algorithm used in the unoptimized case is that 
invoked by the MCK construct {\tt spec\_spr\_nested} -- this essentially performs
BDD-based model checking in a structure in which the worlds are runs of length 
equal to the maximum time relevant to the formula. 

Table~\ref{fig:results-msg} compares the performance of the conditional independence
optimization with this algorithm. The degree of optimization obtained is significant, 
increasing to over four orders of magnitude. 

\begin{table} \centerline{
\begin{tabular}[b]{|c || c | c |   c | c |  }  
\hline 
$n$ & ci (s) & nested(s) & nested/ci \\
\hline 
3& 0.01	& 0.02	& 2\\
4& 0.01	& 0.03	& 3\\
5 & 0.01	& 0.04 & 	4\\
6& 0.01	& 0.06& 	6\\
7& 0.01	& 0.11	& 11\\
8& 0.02	& 0.20	 & 10\\
9 & 0.03	& 0.46& 	15 \\
10 & 0.03	& 1.05	& 35 \\ 
11 & 0.05	& 2.44& 	49\\ 
12 & 0.07	& 5.69	& 81\\
13 & 0.09	& 14.5& 	161\\
14 & 0.12	& 34.77& 	290\\
15 & 0.16	& 89.31& 	558\\
16 & 0.20 & 	360.3& 	1802\\
17 & 0.27	& 1597.91	& 5918\\
\hline 
\end{tabular}}
\caption{\label{fig:results-msg} Message Transmission Protocol, optimized and unoptimized  running times, and speedup ratio  }
\end{table}

\begin{figure} 
\centerline{\includegraphics[height=14cm]{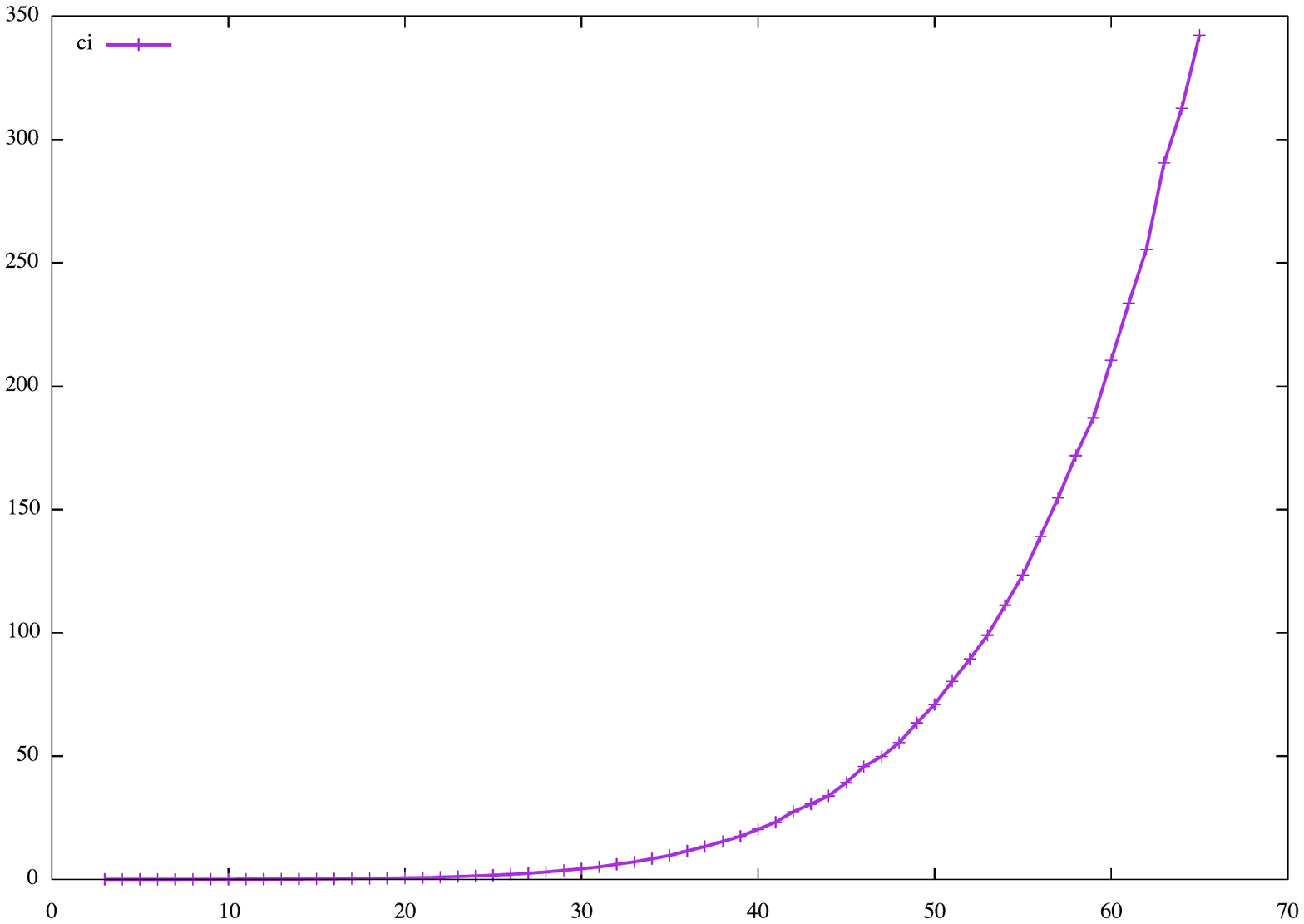}  }

\caption{Message Transmission Protocol, optimized running times
\label{fig:hello-ci-only}} 
\end{figure}

Running the optimization for larger instances, we obtain the 
plot of running times in Figure~\ref{fig:hello-ci-only}.  We again have that 
the optimization enables significantly larger instances to 
be handled in a given amount of time: as many as 65 agents in 
342 seconds, compared to just 16 agents in 360 seconds for the unoptimized version. 

\subsection{Chaum's two-phase protocol} 

The final example we consider is Chaum's two-phase protocol \cite{chaum},  a protocol for anonymous broadcast that 
uses multiple rounds of  the Dining Cryptographers protocol. 
Model checking of this protocol has previously been addressed 
in \cite{BatainehMeyden11}. 

This example scales by both the number of agents and the number of steps of the protocol: 
with $n$ agents, the protocol runs for $O(n)$ steps, and each state is comprised of 
$O(n)$ variables. We check a formula with $O(n)$ variables that says that the first agent 
has a bit {\tt rcvd1} set to true at the end of the protocol iff it knows that 
some other agent is trying to send bit 1. 

The protocol is more complex than the others considered above. 
An initial set of $n$  ``booking'' rounds of the Dining Cryptographers protocol is 
used to anonymously attempt to book one of $n$ slots, 
and this is followed by $n$ ``slot" rounds of the Dining Cryptographers protocol,  
in which an agent who has booked a slot without detecting a collision with another agent's booking, uses that slot to 
attempt to broadcast a message. Because undetected booking collisions remain possible, 
collisions might also be detected in the second phase. 
Because of the complexity of the protocol, this example can 
only be model checked on small instances in reasonable time, even 
with the optimization. Table~\ref{fig:2phase} shows the running times obtained: 
for the unoptimized version, we again used MCK-0.5.1.

\begin{table} \centerline{
\begin{tabular}[b]{|c || c | c |   c |  }  
\hline 
$n$ & ci (s) & xn (s)  \\
\hline 
3  &  0.04 & 0.79\\ 
4  &  0.11 & 96.47\\ 
5 &  0.49 & $>$ 2 hrs \\ 
6 &  2.46 & - \\ 
7 & 12.93 & - \\ 
8  & 155.41 & - \\  
9 & $>$ 2hrs  & - \\
\hline 
\end{tabular}}
\caption{\label{fig:2phase} Chaum's two-phase protocol, optimized and unoptimized  running times  }
\end{table} 

The running time of the unoptimized computation explodes at $n=5$ as we increase the 
number of agents. The optimized computation is significantly less, but also eventually explodes, at $n=9$.
Thus, the optimization has doubled the size of the problem that can be handled in reasonable time.

\section{Related Work and Conclusion} \label{sec:concl} 

We conclude with a discussion of some related work and future directions. 

Wilson and Mengin \cite{WM01} have previously related modal logic to valuation algebra, 
but their definition requires that the marginalization of a Kripke structure have exactly the 
same set of worlds and equivalence relation, and merely restricts the assignment at 
each world, so their approach does not give the optimization that we have developed, 
and a model checking approach based on it would be less efficient than that developed in the 
present paper. They do not discuss conditional independence, which is a key part of our approach. 

Also related are probabilistic programs, a type of program containing probabilistic choice statements, that
sample from a specified distribution. The semantics of such programs is that they generate a 
probability distribution over the outputs. 
These programs may contain statements of the form $observe(\phi)$ where $\phi$ is a boolean condition: 
these are interpreted as conditioning the distribution constructed to that point on the condition $\phi$. 
Hur et al.~\cite{HurNRS14} develop an approach to slicing probabilistic programs based on a static analysis that
incorporates ideas from the Bayesian net literature. 
There are several  differences between probabilistic programs and our work in this paper. 
One is that we deal with discrete knowledge rather than probability -- in general, 
this makes our model checking problem more tractable.  We also
reason about all possible sequences of observations, rather 
than one particular sequence of observations. 
Additionally, we allow observations by multiple agents rather than just one. 
Finally, via knowledge operators, we have a locus of reference to observations in our framework that is located in 
formulas rather than inside the program -- this enables us to ask multiple questions about a program 
without changing the code, whereas in probabilistic programs, one would need to handle this
by multiple distinct modifications of the code. 

The results of the present paper concern formulas that refer (directly and through knowledge operators) only
to a specific time. Our approach, however, can be easily extended by means of a straightforward  
transformation to formulas that talk about multiple time points, and we intend to implement
this extension in future work. 

The technique we have developed can also be extended to deal with multi-agent models 
based on programs taking probabilistic transitions, which MCK already supports. 
Formulas in this extension would include operators that talk about an 
agent's subjective probability, given what it has observed. 

Other extensions we intend to implement are to enrich the range of knowledge semantics
beyond the synchronous perfect recall semantics treated in this paper: essentially 
the same techniques will apply to the clock semantics (in which an agent's knowledge is based on 
just its current observation and the current time). The observational semantics, in which the agent's knowledge
is based just on its current observation, will be more challenging, since it is asynchronous, and 
knowledge formulas may refer to times arbitrarily far into the future. 

Finally, whereas the present paper concentrated on straightline programs, we
intend to extend to a richer protocol format, including conditionals. 
In general, this extension will diminish the power of the 
equality optimization, and result in an increased set of dependencies in which 
many variables become dependent on a variable representing the program counter. 
However, some extensions will be more managable, e.g., one that requires conditionals and loops to be balanced
in their timing (a restriction already used in work on computer security 
to avoid unwanted information leakages).

\bibliographystyle{plain} 
\bibliography{condind}

\begin{full} 
\newpage

\appendix

\section{Details of Experiments} 
\label{appendix} 

In this appendix we provide further details on the examples considered
in our experiments. 

The MCK scripts below start with a declaration of global variables, 
followed by the ``\verb+init_cond+'' construct which gives a boolean
formula describing the initial states of the model.  
This is followed by the declaration of the agents in model.
Each declaration names the agent, gives the name of the 
protocol it runs (in quotes), followed by the binding of the 
parameters of this protocol to environment variables.  
The protocols are listed last in the script. 
All of the agent protocols are straightline. 
The bracket notation  ``\verb+<| ... |>+'' delimits atomic actions. The contents of these 
brackets are a sequence of assignments that execute atomically, without consuming time. 

The intuitive operational semantics of the scripts is that at time $n$, the agents activate the next such action in the sequence. 
These actions are performed in the order of the agents listed, 
followed by the sequence of assignments in the ``transitions" clause, which 
intuitively, describes events that happen in the environment at each step. 
The resulting state is then taken to be the state at time $n+1$. 

Specifications are listed using the construct ``\verb+spec_spr+''. Here
the ``\verb+spr+'' indicates that we are using a synchronous perfect recall semantics for knowledge.

\subsection{Dining Cryptographers} 

The Dining Cryptographers protocol has already been discussed in the 
body of the paper, we provide the code for the 3-agent instance below. 
This example is generalized to larger instances by increasing the 
number $n$ of agents: each running the protocol given. 
The protocol can be run using any connected 
network, but  we use a ring network, with agent $C_i$ sharing coins
with agent $C_{i-1\mod n}$ to the left, and agent agent $C_{i+1\mod n}$ to the right. 
The query is stated in terms of the knowledge of agent $C_0$, and 
says that this agent either knows that nobody paid, 
knows that it paid itself, or knows that one of the other $n-1$ agents 
paid, but does not know which. 

{\small
\newpage 
\begin{verbatim} 
paid : Bool[3]
chan : Bool[3]
said : Bool[3]

init_cond = 
((neg paid[1]) /\ (neg paid[2])) \/
((neg paid[0]) /\ (neg paid[2])) \/
((neg paid[0]) /\ (neg paid[1])) 


agent C0 "dc_agent_protocol" (paid[0], chan[0], chan[1], said, said[0])

agent C1 "dc_agent_protocol" (paid[1], chan[1], chan[2], said, said[1])

agent C2 "dc_agent_protocol" (paid[2], chan[2], chan[0], said, said[2])

spec_spr_ci = X 3 (Knows C0 ((neg paid[0])  /\ (neg paid[1]) /\ (neg paid[2]))) \/
   (Knows C0 (paid[0]))  \/
   (Knows C0 ( False \/ paid[1]\/ paid[2])  /\ 
      (neg Knows C0 (neg paid[1]))/\ (neg Knows C0 (neg paid[2])))


protocol "dc_agent_protocol"
(
  paid : observable Bool,
  chan_left : Bool,
  chan_right : Bool,
  said : observable Bool[], -- the broadcast variables.
  say : Bool 
)

coin_left : observable Bool
coin_right : observable Bool

begin
  <| chan_right := coin_right |>;
  <| coin_left := chan_left |>;
  <| say := coin_left xor coin_right xor paid |>  ; 
  skip 
end
\end{verbatim} 
}

\subsection{One Time Pad} 

The one-time pad is a shared secret key to be used just once. It is known
to give perfect encryption under this assumption. We model a system in 
which Alice has a message (a boolean string) to transmit to Bob. 
She is assumed to share a one-time pad, another boolean string of the same 
length as the message, with Bob. Alice encrypts her string by bitwise exclusive-or 
with the one-time pad, and sends the resulting encrypted bits via
a channel that is observed by an eavesdropper Eve. 

We parameterize this model by the length $n$ of the strings. 
The case of $n=3$ is shown below. We consider a query that
states that at the end of the transmission, Eve does not know 
the value of the first bit of Alice's message.

{\small 
\begin{verbatim} 
-- The 'secret' one-time-pad shared between Alice and Bob.
one_time_pad : Bool[3]
-- The communications channel.
channel : Bool

agent Alice "sender"      (one_time_pad, channel)
agent Bob   "receiver"   (one_time_pad, channel)
agent Eve   "eavesdropper" (channel)

spec_spr  = 
 X 6  ((neg (Knows Eve Alice.message[0])) /\ (neg (Knows Eve (neg Alice.message[0]))))

-- Alice's protocol.
protocol "sender" (otp : Bool[3], chan : Bool)

message : Bool[3]
bit : Bool 

begin
  <| bit := otp[0] |>; <| chan := message[0] xor bit |>; 
  <| bit := otp[1] |>; <| chan := message[1] xor bit |>; 
  <| bit := otp[2] |>; <| chan := message[2] xor bit |>
end

-- Bob's protocol.
protocol "receiver" (otp : observable Bool[3], chan : observable Bool)
begin
skip; skip; skip; 
skip; skip; skip
end

-- Eve's protocol.
protocol "eavesdropper" (chan : observable Bool)

begin
skip; skip;  skip; 
skip; skip; skip
end
\end{verbatim} }

\subsection{Rivest's Oblivious Transfer Protocol} 

Rivests Oblivious Transfer protocol \cite{rivest99} enables a receiver Bob
to obtain exactly one of two distinct messages $m_0$, $m_1$
possessed by a sender Alice, without Alice learning
which message Bob chose to receive. That is, Bob makes a
choice $c$, and, at the end of the protocol, knows message $m_c$, but 
not the other message $m_{\overline{c}}$, without Alice learning
the value of $c$. 

An MCK model of the protocol in the case where the length of the message 
is 3 is given. The protocol requires that Alice and Bob start with 
some correlated randomness  which can be provided by a
trusted third party who does not need to be online during the 
running of the protocol. This trusted third party 
provides Alice with two random strings $r_0,r_1$, 
and Bob with a random bit $d$ and the string $r_d$. 
We do not model this third party explicitly, but start a the
state where Alice and Bob have received this information. 
(As with the Dining Cryptographers protocols above, we
model random choices as nondeterministic choices.) 

Bob and Alice then exchange some messages computed from the initial information. 
Bob first sends a bit $e$, and Alice responds with two strings $f_0,f_1$
that encode $m_0$ and $m_1$. Bob is then able to compute his desired message $m_c$
in the third step of the protocol. 

For this protocol, we scale our experiments by the length $n$ of the messages $m_0,m_1$
(the protocol always runs in 3 steps, but the last step is a local computation by Bob, so does not
affect the agent's knowledge). We consider two formulas: 
\begin{enumerate} 

\item  The first ``[Single]''  says that if Bob chose to receive $m_1$, then he does not learn the first bit of $m_0$. 
Because the protocol effectively operates independently on the bits of the various messages, we 
expect that the dependency analysis will detect this independence and give a significant speedup
as the size of the messages increase. 

\item  The first ``[Any]'' says that if Bob chose to receive $m_1$, then he does not learn any bit of $m_0$. 
This involves $n$ variables, so it is not immediately clear whether we should expect any optimization. 
\end{enumerate}

{\small
\begin{verbatim} 
-- Alice's messages 
m0: Bool[3]
m1: Bool[3]

-- A variable used by Bob to store the message received
mc: Bool[3]

-- initial randomness  
r0 : Bool[3]
r1 : Bool[3]
rd : Bool[3]
d : Bool

f0 : Bool[3]
f1 : Bool[3]
e : Bool
c: Bool

init_cond = 
-- Message rd is determined from r0,r1 and d.  
( neg d => ((r0[0] <=> rd[0])  /\ (r0[1] <=> rd[1])  /\ (r0[2] <=> rd[2]) )) /\ 
        (d => ((r1[0] <=> rd[0])  /\ (r1[1] <=> rd[1])  /\ (r1[2] <=> rd[2]) )) /\ 
-- The random strings are distinct. 
( neg (r0[0] <=> r1[0])  \/ neg (r0[1] <=> r1[1])  \/ neg (r0[2] <=> r1[2]) ) /\ 
-- The messages m0, m1 are distinct. 
( neg (m0[0] <=> m1[0])  \/ neg (m0[1] <=> m1[1])  \/ neg (m0[2] <=> m1[2]) )

agent Alice "alice" (r0, r1, m0, m1, f0, f1, e)
agent Bob "bob" (e, rd, d, c, f0, f1, mc)


spec_spr = 
"[Any]: after two steps, Bob does not know the value of any bit of m0" 
X 2  ( c => (neg (Knows Bob m0[0]) /\ neg (Knows Bob neg m0[0]) /\ 
                   neg (Knows Bob m0[1]) /\ neg (Knows Bob neg m0[1]) /\ 
                   neg (Knows Bob m0[2]) /\ neg (Knows Bob neg m0[2]))) 

spec_spr = 
"[Single]: after two steps, Bob does not know the value of the first bit of m0" 
X 2  (neg (Knows Bob m0[0]) /\ neg (Knows Bob neg m0[0])) 

spec_spr  = "[Alice] Alice does not learn Bob's choice: " 
X 3 ( (neg Knows Alice c) /\ (neg Knows Alice neg c ) ) 


protocol "alice" (r0 : observable Bool[3], r1: observable Bool[3],
                  m0 : observable Bool[3], m1: observable Bool[3], 
                  f0 : observable Bool[3], f1: observable Bool[3], 
                  e: observable Bool)

begin
  skip; 
  <|
     f0[0]:= ( (neg e) /\ (m0[0] xor r0[0])) \/ (e /\ (m0[0] xor r1[0])) ; 
     f1[0]:= ( (neg e) /\ (m1[0] xor r1[0])) \/ (e /\ (m1[0] xor r0[0])) ;
      
     f0[1]:= ( (neg e) /\ (m0[1] xor r0[1])) \/ (e /\ (m0[1] xor r1[1])) ; 
     f1[1]:= ( (neg e) /\ (m1[1] xor r1[1])) \/ (e /\ (m1[1] xor r0[1])) ; 
     
     f0[2]:= ( (neg e) /\ (m0[2] xor r0[2])) \/ (e /\ (m0[2] xor r1[2])) ; 
     f1[2]:= ( (neg e) /\ (m1[2] xor r1[2])) \/ (e /\ (m1[2] xor r0[2])) 
  |>; 
  skip 
end


protocol "bob" (e: Bool, 
       rd: observable Bool[3], d: observable Bool, c: observable Bool, 
       f0: observable Bool[3], f1: observable Bool[3], mc: observable Bool[3])
begin
 <| e:= d xor c |>; 
 skip; 
 <| 
   mc[0]:= ((neg c) /\ (f0[0] xor rd[0])) \/ (c /\ (f1[0] xor rd[0])) ; 
   mc[1]:= ((neg c) /\ (f0[1] xor rd[1])) \/ (c /\ (f1[1] xor rd[1])) ; 
   mc[2]:= ((neg c) /\ (f0[2] xor rd[2])) \/ (c /\ (f1[2] xor rd[2]))  
 |> 
end 

\end{verbatim} 
}

\subsection{Message Transmission with Uncertain Delay}  

This example models a scenario where an agent Alice
sends a message $x$ to Bob through a channel with a bounded delay. 
The example is parameterized by the maximum length $n$ of the delay. 
The instance with $n=3$ is shown. 

Initially, all variables except Alice's local variable $x$ and the array $\mathit{delay}$ are 0. 
At time $0$, Alice writes the message $x$ to a buffer, and 
at time $1$, Alice sets a bit to $\mathit{True}$ to 
start the transmission. The message 
is delivered as soon as the value $\mathit{delay}[0]$ is  $\mathit{True}$. 
Here $\mathit{delay}$ is an array of length $n$, which initially has a random 
value. At each step, the values in the array shift to the left by one position, 
and the final value is set to be $\mathit{False}$. 
Thus, $\mathit{delay}[0]$ becomes $F$ by time $n$ at the latest, and the message 
is guaranteed to be delivered by time $n+1$. 

In this example, we consider a query that involves nested knowledge. 
It is the case that Alice considers it possible up 
to time $n+1$ that the message has not yet been delivered, but
by time  $n+1$ Alice knows that the message must have been delivered. 
In fact, it is common knowledge at time $n+1$ that the message has been delivered. 
We consider a query that says that Alice Knows that Bob Knows that Alice 
Knows that Bob Knows  that Alice Knows the message has been received by Bob.

{\small 
\begin{verbatim} 
delay : Bool[3]
outA : Bool 
sentA : Bool

inB : Bool 
rcdB : Bool 

init_cond = neg (sentA \/ outA \/ inB \/ rcdB) 

agent Alice "sender" (outA, sentA)
agent Bob  "receiver" (inB, rcdB)

transitions 
begin 
-- delay[0] captures whether transmission is delayed in the current step 
-- if there is no delay and Alice has sent, then Bob receives 

rcdB := rcdB \/ (neg delay[0] /\ sentA ); 
inB := (neg delay[0] /\ sentA /\ outA) \/ ((delay[0] \/ neg sentA) /\ inB); 

-- delay starts out random, and shifts from right to left 

delay[0] := delay[1] ; 
delay[1] := delay[2];
delay[2] := False
 end 

spec_spr = X 4 Knows Alice (Knows Bob (Knows Alice (Knows Bob (Knows Alice rcdB ))))

-- Alice's protocol.
protocol "sender" (chan : Bool, sent : Bool )

x: Bool 
 
begin
<| chan := x |> ; 
<| sent := True |> ; 
skip; skip; skip 
end



-- Bob's protocol.
protocol "receiver" (chanin: observable Bool, rcd: observable Bool)

begin
skip; skip; skip; skip 
end
\end{verbatim} 
}

\subsection{Two-Phase Protocol} 

In the Dining Cryptographers protocol, it is assumed
that at most one agent wishes to communicate the message that they paid. 
The two-phase protocol, also from \cite{chaum}, is an application of the 
basic Dining Cryptographers protocol, in which 
multiple rounds of the basic dining cryptographers protocol
are used to allow anonymous broadcast in a setting
where multiple agents may have a message to send.  

We model an abstraction of this protocol, in which
the message of each agent consists of a single bit, 
and each application of Dining Cryptographers scheme
is represented by taking the exclusive-or of the bits
contributed to the round by each of the agents.
It is shown in \cite{BatainehMeyden11} that this 
abstraction is sound for verification  of epistemic properties. 

The protocol operates in two phases. The second phase consists of some number $m$ of rounds 
of the basic Dining Cryptographers protocol, which some
agent may use to anonymously broadcast a message. Of 
course, to remain anonymous, the slot in which it 
chooses to broadcast must not be predictable, so 
the agent must choose its slot randomly. This creates the risk of 
collisions, in which two agents broadcast in the same slot. 
This is problematic:  the Dining Cryptographers protocol assumes that 
at most one agent is sending a message. 

To decrease the risk of a collision, in the first phase, any agent with a message to broadcast
attempts to book one of the broadcast slots. To maintain its  anonymity, the booking of a 
slot must also be done anonymously, and the Dining Cryptographers protocol is also used by the agent 
to announce that it wishes to use its chosen slot. Thus, the first phase also consists of $m$ rounds 
of the Dining Cryptographers protocol. 

Some collisions can be detected during the first round -- in particular, if an even 
number of agents attempt to book the same slot, they will detect the collision 
from the outcome of that booking round.  
There remains the possibility of undetected collisions --- the idea of the protocol
is that these will be detected during the transmission round, and that in this
case, the agent will make another attempt to transmit in a later run of the protocol.  

This makes the precise conditions under which it is known that a message has been successfully transmitted using the 
two-phase protocol quite subtle. For a longer discussion of the subtleties, see \cite{BatainehMeyden11}. 
We  focus on one of the simpler queries from that analysis, concerning whether 
an agent knows that some other agent is transmitting a particular bit.

In the script,  the boolean array \verb+slotsC+$i$ is used to represent which 
slot, if any, agent \verb+C+$i$ has selected for its transmission;
\verb+slotsC+$i[0]$ represents that the agent has nothing to transmit. 
The initial condition states that at exactly one of \verb+slotsC+$i[j]$ holds, for $j=0\ldots n$.  
The variables \verb+rr+$i$ are used to capture the results of the reservation round $i$
during the first phase. These values are then used during the corresponding round of the 
transmission phase to determine whether a value has been received in that round. 
Variable \verb+rcvd+$X$ for $X\in \{0,1\}$ is used to represent that a bit $X$ has been received. 

This example scales by increasing the number of agents or increasing the number of available slots. 
We have experimented with two versions, one where the number of slots is fixed at 5,
and we vary the number of agents, and another, where we vary the number of agents 
and keep this equal to the number of slots. 

The query that we consider states that the variable \verb+rcvd1+ corresponds 
precisely to the agent knowing that some agent is transmitting the bit $1$. 
(We verify this for agent C0, it holds for the others by symmetry.)

{\small 
\begin{verbatim} 
 type Slot = {0..3}
slotsC0 : Bool[Slot] 
slotsC1 : Bool[Slot] 
slotsC2 : Bool[Slot] 
say : Bool[3]
round_result : Bool

 init_cond = 
(neg (slotsC0[0] /\ slotsC0[1])) /\ (neg (slotsC0[0] /\ slotsC0[2])) /\  
(neg (slotsC0[0] /\ slotsC0[3])) /\ (neg (slotsC0[1] /\ slotsC0[2])) /\ 
(neg (slotsC0[1] /\ slotsC0[3])) /\ (neg (slotsC0[2] /\ slotsC0[3])) /\ 
(neg (slotsC1[0] /\ slotsC1[1])) /\ (neg (slotsC1[0] /\ slotsC1[2])) /\ 
(neg (slotsC1[0] /\ slotsC1[3])) /\ (neg (slotsC1[1] /\ slotsC1[2])) /\ 
(neg (slotsC1[1] /\ slotsC1[3])) /\ (neg (slotsC1[2] /\ slotsC1[3])) /\ 
(neg (slotsC2[0] /\ slotsC2[1])) /\ (neg (slotsC2[0] /\ slotsC2[2])) /\ 
(neg (slotsC2[0] /\ slotsC2[3])) /\ (neg (slotsC2[1] /\ slotsC2[2])) /\ 
(neg (slotsC2[1] /\ slotsC2[3])) /\ (neg (slotsC2[2] /\ slotsC2[3])) /\ 
 (slotsC0[0] \/ slotsC0[1] \/ slotsC0[2] \/ slotsC0[3])  /\ 
 (slotsC1[0] \/ slotsC1[1] \/ slotsC1[2] \/ slotsC1[3])  /\ 
 (slotsC2[0] \/ slotsC2[1] \/ slotsC2[2] \/ slotsC2[3])  

agent C0 "twophase_protocol" (slotsC0, say[0], round_result) 
agent C1 "twophase_protocol" (slotsC1, say[1], round_result) 
agent C2 "twophase_protocol" (slotsC2, say[2], round_result) 

transitions
begin
round_result := say[0] xor say[1] xor say[2] 
end

 -- rcvdX = I know someone else is sending X

spec_spr = X 13 C0.rcvd1 <=> 
Knows C0 ((neg slotsC1[0] /\ C1.message) \/ (neg slotsC2[0] /\ C2.message))

protocol "twophase_protocol"
(
  slot_request: observable Bool[],
  say : Bool, 
  round_result: observable Bool
)

-- the following variables are initialised nondeterministically:

-- the message the agent sends, if any 
message : observable Bool

-- the result for each DC round 
-- rri = message received in booking round i 
rr1 : Bool 
rr2 : Bool 
rr3 : Bool 

--  rcvdX = I know a message X has been sent by someone else
rcvd0 :  Bool
rcvd1 :  Bool



begin
-- reservation phase
-- time 0 
<| say := slot_request[1] |>; 

<| rr1 := round_result |>;  

<| say := slot_request[2] |>; 

<| rr2 := round_result |>;  

<| say := slot_request[3] |>; 

<| rr3 := round_result |>;  

--initialize rcvd vars
<| rcvd0:= False ; rcvd1 := False |>; 

 -- Sending phase: 
<| say := (slot_request[1] /\ rr1 /\ message ) |> ; 

<| 
    rcvd1 := rcvd1 \/ (neg slot_request[1] /\ rr1 /\ round_result) \/ 
             (slot_request[1] /\ rr1 /\ (message xor round_result));   

    rcvd0 := rcvd0 \/ (neg slot_request[1] /\ rr1 /\ neg round_result) \/ 
             (slot_request[1] /\ rr1 /\ (message xor round_result))   
|> ; 

<| say := (slot_request[2] /\ rr2 /\ message ) |> ; 

<| 
    rcvd1 := rcvd1 \/ (neg slot_request[2] /\ rr2 /\ round_result) \/ 
             (slot_request[2] /\ rr2 /\ (message xor round_result));   

    rcvd0 := rcvd0 \/ (neg slot_request[2] /\ rr2 /\ neg round_result) \/ 
             (slot_request[2] /\ rr2 /\ (message xor round_result))   
|> ; 

<| say := (slot_request[3] /\ rr3 /\ message ) |> ; 

<| 
    rcvd1 := rcvd1 \/ (neg slot_request[3] /\ rr3 /\ round_result) \/ 
             (slot_request[3] /\ rr3 /\ (message xor round_result));   

    rcvd0 := rcvd0 \/ (neg slot_request[3] /\ rr3 /\ neg round_result) \/ 
             (slot_request[3] /\ rr3 /\ (message xor round_result))   
|> 
 
end
\end{verbatim} 
}

\end{full} 
\end{document}